\documentclass[fleqn,usenatbib]{mnras}

\usepackage{newtxtext,newtxmath}

\usepackage[T1]{fontenc}
\usepackage{ae,aecompl}

\usepackage{graphicx}	
\usepackage{amsmath}	
\usepackage{amssymb}	

\title[Exterior Test Particle]{Secular Dynamics 
of an Exterior Test Particle:\\The Inverse Kozai and Other Eccentricity-Inclination Resonances}

\author[Vinson \& Chiang]{
Benjamin R.~Vinson$^{1}$\thanks{E-mail: ben\_vinson@berkeley.edu},
Eugene Chiang$^{2,3}$\thanks{E-mail: echiang@astro.berkeley.edu}
\\
$^{1}$Department of Physics, 366 LeConte Hall, University of California, Berkeley, 94720, USA\\
$^{2}$Department of Astronomy, 501 Campbell Hall, University of California, Berkeley 94720-3411, USA\\
$^{3}$Department of Earth and Planetary Science, 307 McCone Hall, University of California, Berkeley 94720, USA
}

\pubyear{2017}

\begin{document}
\label{firstpage}
\pagerange{\pageref{firstpage}--\pageref{lastpage}}
\maketitle

\begin{abstract}
The behavior of an interior test particle in the secular 
three-body problem has been studied extensively. A
well-known feature is the Lidov-Kozai resonance 
in which the test particle's argument of periastron librates about
$\pm 90^\circ$ and large oscillations in eccentricity and
inclination are possible.
Less explored is the inverse problem: the dynamics of an exterior
test particle and an interior perturber.
We survey numerically the inverse secular problem,
expanding the potential to hexadecapolar order 
and correcting an error in the published expansion.
Four secular resonances are uncovered that persist in
full $N$-body treatments (in what follows,
$\varpi$ and $\Omega$
are the longitudes of periapse and of ascending node,
$\omega$ is the argument of periapse, and subscripts 1 and
2 refer to the inner perturber and the outer test particle):
(i) an orbit-flipping
quadrupole resonance requiring a non-zero perturber eccentricity
$e_1$,
in which 
$\Omega_2-\varpi_1$ librates about $\pm 90^\circ$;
(ii) a hexadecapolar resonance (the ``inverse Kozai'' resonance)
for perturbers that are circular or nearly so
and inclined by $I \simeq 63^\circ/117^\circ$,
in which 
$\omega_2$ librates about $\pm 90^\circ$
and which can vary the particle eccentricity by $\Delta e_2 \simeq 0.2$
and lead to orbit crossing;
(iii) an octopole ``apse-aligned'' resonance at
$I \simeq 46^\circ/107^\circ$ wherein 
$\varpi_2 - \varpi_1$
librates about $0^\circ$
and $\Delta e_2$ grows with $e_1$; and
(iv) an octopole resonance at $I \simeq 73^\circ/134^\circ$ wherein 
$\varpi_2 + \varpi_1 - 2 \Omega_2$
librates about
$0^\circ$ and $\Delta e_2$ can be as large as 0.3 for small
but non-zero $e_1$. 
Qualitatively, the more eccentric the perturber, the more the particle's eccentricity
and inclination vary; also, more polar orbits are more chaotic.
Our solutions to the inverse problem have
potential application to the Kuiper belt and
debris disks, circumbinary planets, and hierarchical stellar systems.
\end{abstract}

\begin{keywords}
celestial mechanics -- binaries: general -- planets and satellites: dynamical evolution and stability -- Kuiper belt: general
\end{keywords}



\section{Introduction} \label{sec:intro}

In the restricted three-body problem, the Lidov-Kozai resonance
provides a way
for an external perturber to torque test particle orbits
to high eccentricity and inclination \citep{lidov62,kozai62}.
When the perturber's orbit is circular ($e_2 = 0$),
and when the inclination $I$
between the test particle's orbit and the perturber's exceeds
$\arccos \sqrt{3/5} \simeq 39^\circ$, the test particle's
argument of periastron $\omega_1$ can librate (oscillate) about either
90$^\circ$ or 270$^\circ$: these are the fixed points of the
Lidov-Kozai (``Kozai'' for short) resonance.\footnote{Throughout this
paper, subscript ``1'' denotes the interior body and subscript ``2''
denotes the exterior body; by definition, the orbital semimajor axes
$a_1<a_2$. These subscripts apply regardless of whether the body is a test
particle or a perturber.}
The larger the libration amplitude, 
the greater the eccentricity
variations. For circular perturbers, the test particle
eccentricity $e_1$ can cycle between 0 and 1 as the inclination $I$
cycles between $90^\circ$ and $39^\circ$;\footnote{There is also a
retrograde branch for the standard Kozai resonance in which
$I$ cycles between $90^\circ$ and $141^\circ$. In this paper we will
encounter several resonances for which retrograde fixed points
are paired with prograde fixed points,
but will sometimes focus on the prograde branches for
simplicity.} $e_1$ and $I$ seesaw
to conserve $J_{1z} \propto \sqrt{1-e_1^2} \cos I$,
the test particle's vector angular momentum projected onto
the perturber's orbit normal. 
For eccentric external perturbers ($e_2 \neq 0$),
the gravitational potential is no longer
axisymmetric, and the test particle's $J_{1z}$ is now free to vary,
which it can do with a vengeance: the test particle can start from a
nearly coplanar, prograde orbit ($J_{1z} > 0$) and ``flip'' to being
retrograde ($J_{1z} < 0$; e.g., \citealt{lithwick11} and
\citealt{katz11}).
The large eccentricities and inclinations
accessed by the Kozai mechanism have found application in numerous
settings: enabling Jupiter to send comets onto sun-grazing trajectories
(e.g., \citealt{bailey92});
delineating regions of orbital stability for
planetary satellites perturbed by exterior satellites and the Sun
(e.g., \citealt{carruba02};
\citealt{nesvorny03}; \citealt{tremaine14});
merging compact object binaries in triple systems
(e.g., \citealt{kushnir13}; \citealt{silsbee17});
and explaining the orbits of eccentric or short-period
extrasolar planets (e.g., \citealt{wu03}), including
warm Jupiters (\citealt{dawson14}), and hot Jupiters
with their large spin-orbit obliquities \citep{naoz11}.
See \cite{naoz16} for a review.

The Kozai resonance is a secular effect (i.e., it does not depend on
orbital longitudes, which are time-averaged away from the equations
of motion) that applies to an interior
test particle perturbed by an exterior body.
Curiously, the ``inverse'' secular problem---an
exterior test particle and an interior perturber---does not seem to
have received as much attention as the conventional problem.
\citet{gallardo12} studied the inverse problem in the context
of Kuiper belt objects perturbed by Neptune and the other giant planets,
idealizing the latter as occupying circular ($e_1=0$) and coplanar orbits,
and expanding the disturbing function (perturbation potential)
to hexadecapolar order in $\alpha \equiv a_1/a_2$.
They discovered an analogous Kozai resonance
in which $\omega_2$ librates
about either $+90^\circ$ or $-90^\circ$, 
when $I \simeq \arccos \sqrt{1/5} \simeq 63^\circ$
(see also \citealt{tremaine14}). Eccentricity variations are
stronger inside this ``inverse Kozai'' or $\omega_2$ resonance
than outside. \citet{thomas96}
also assumed the solar system giant planets to be on circular
coplanar orbits, dispensing with a multipole expansion and numerically
computing secular Hamiltonian level curves for an exterior
test particle. For $a_2 = 45$ AU, just outside Neptune's orbit,
resonances appear at high $I$ and $e_2$ that are
centered on $\omega_2 = 0^\circ$ and $180^\circ$.

\citet{naoz17} also studied the inverse
problem, expanding the potential to octopolar order and considering non-zero
$e_1$. Orbit flipping was found to be possible via a quadrupole-level
resonance 
that exists only when $e_1 \neq 0$ and 
for which $\Omega_2-\varpi_1$ librates about either $+90^\circ$ or $-90^\circ$.
Here $\Omega$ and $\varpi$ are the longitudes of ascending node
and of periastron, respectively.
As $e_1$ increases, the minimum $I$ at which orbits can flip decreases.
All of this inclination behavior obtains at the quadrupole level;
the $\Omega_2-\varpi_1$ resonance was also uncovered by
\citet{verrier09} and studied analytically by \citet{farago10}. 
Octopole terms were shown by \citet{naoz17} 
to enable test particles to alternate between
one libration center ($\Omega_2-\varpi_1 = +90^\circ$) and another
($\Omega_2-\varpi_1 = -90^\circ$), modulating the inclination evolution
and introducing chaos, particularly at high $e_1$.

In this paper we explore more systematically 
the inverse problem, expanding the perturbation potential
to hexadecapolar order 
and considering non-zero $e_1$. In addition to studying more closely
the hexadecapolar inverse Kozai resonance found by \citet{gallardo12}
and how it alters when $e_1$ increases, we will uncover
strong, octopolar resonances not identified by \citet{naoz17}.
By comparison with the latter work, we focus more on the test
particle's eccentricity variations than on its inclination variations.
We are interested, for example, in identifying dynamical channels
that can connect planetary systems with more distant reservoirs
of minor bodies, e.g., the ``extended scattered'' or ``detached''
Kuiper belt (e.g., \citealt{sheppard16}), or the Oort cloud (e.g.,
\citealt{silsbee16}). Another application is to extrasolar debris disks,
some of whose scattered light morphologies appear sculpted by
eccentric perturbers (e.g., \citealt{lee16}).
We seek to extend such models to large mutual inclinations
(e.g., \citealt{verrier08}; \citealt{pearce14}; \citealt{nesvold16}; \citealt{zanardi17}).

This paper is organized as follows. In Section \ref{sec:disturb}
we write down the secular disturbing function of an interior
perturber to hexadecapolar order in $\alpha = a_1/a_2$. There we
also fix certain parameters (masses and semimajor axes; e.g., $\alpha = 0.2$)
for our subsequent numerical survey.
Results for a circular perturber are given in Section \ref{sec:circular} 
and for an eccentric perturber in Section \ref{sec:eccentric}, 
with representative secular integrations tested
against $N$-body integrations. We wrap up in Section \ref{sec:conclude}.

\section{Secular Disturbing Function for Exterior Test Particle}
\label{sec:disturb}

The disturbing function of
\citet[][hereafter Y03]{yokoyama03} is
for the conventional problem: an interior test particle
(satellite of Jupiter)
perturbed by an exterior body (the Sun). 
We may adapt their $R$ to our inverse problem
of an exterior test particle perturbed by
an interior body by a suitable reassignment of variables.
This reassignment is straightforward because the disturbing function
is proportional to $1/\Delta$, where $\Delta$ is the absolute magnitude
of the distance between 
the test particle and the perturber (Y03's equation 2, with the indirect
term omitted because that term vanishes after secular averaging).
This distance $\Delta$ is obviously the same between the conventional and inverse
problems, and so the Legendre polynomial expansion of $1/\Delta$
performed by Y03 for their problem holds just as well for ours.

The change of variables begins with a simple replacing of subscripts.
We replace Y03's subscript $\odot$ (representing the Sun, their exterior
perturber) with ``2'' (our exterior test particle). For their unsubscripted
variables (describing their interior test particle),
we add a subscript ``1'' (for
our interior perturber). Thus we have:
\begin{align}
a &\rightarrow a_1 \label{eq:a1}\\
a_\odot &\rightarrow a_2 \\
e &\rightarrow e_1 \\
e_\odot &\rightarrow e_2 \, ,
\end{align}
where $a$ is the semimajor axis and $e$ is the eccentricity.
Since we are interested in the inverse problem
of an interior perturber, we replace their perturber mass $M_\odot$
with our perturber mass:
\begin{equation}
M_\odot \rightarrow m_1 \,.
\end{equation}
Their inclination $I$ is the mutual inclination between the interior
and exterior orbits; we leave it as is.

Mapping of the remaining angular variables requires more care.
Y03's equations (6) and (7) take the reference plane to coincide with
the orbit of the exterior body---this is the invariable plane
when the interior body is a test particle. We want the reference
plane to coincide instead
with the orbit of the interior body (the invariable plane
for our inverse problem). To convert to the new reference
plane, we use the relation 
\begin{equation}
\Omega_1 - \Omega_2 = \pi
\end{equation}
for longitude of ascending node $\Omega$,
valid whenever the reference plane is the invariable plane
for arbitrary masses 1 and 2
(the vector pole of the invariable plane is co-planar
with the orbit normals of bodies 1 and 2, and lies between them). 
We therefore map Y03's $\Omega$ to
\begin{equation} \label{eq:Omega}
\Omega \rightarrow \Omega_1 \rightarrow \Omega_2 + \pi \,,
\end{equation}
and their argument of periastron $\omega$ to
\begin{equation} \label{eq:omega}
\omega \equiv (\varpi - \Omega) \rightarrow (\varpi_1 - \Omega_1) \rightarrow (\varpi_1 -
\pi - \Omega_2)
\end{equation}
where $\varpi$ is the longitude of periastron.
Although $\varpi_1$ remains meaningful in our new reference plane,
$\Omega_1$ and $\omega_1$ are no longer meaningful, and are
swapped out using (\ref{eq:Omega}) and (\ref{eq:omega}).
Finally
\begin{equation} \label{eq:varpi2}
\varpi_\odot \rightarrow \varpi_2 \,.
\end{equation}

Armed with (\ref{eq:a1})--(\ref{eq:varpi2}), we 
re-write equations (6)--(8) of
Y03 to establish the secular
disturbing function $R$ for an exterior test particle
perturbed by an interior body of mass $m_1$, expanded to hexadecapole
order:
\begin{align}
b_1 &= -(5/2)e_1 - (15/8)e_1^3 \\
b_2 &= -(35/8)e_1^3 \\
c_1 &= Gm_1a_1^2 \frac{1}{a_2^3 (1-e_2^2)^{3/2}} \\
c_2 &= Gm_1a_1^3 \frac{e_2}{a_2^4 (1-e_2^2)^{5/2}} \\
c_3 &= Gm_1a_1^4 \frac{1}{a_2^5 (1-e_2^2)^{7/2}} \\
d_1^\ast &= 1 + (15/8)e_1^2 + (45/64)e_1^4 \label{eq:d1star} \\
d_2 &= (21/8) e_1^2 (2 + e_1^2) \\
d_3 & = (63/8)e_1^4 \\
cI &\equiv \cos I \\
sI &\equiv \sin I \\
R_2 &= \frac{1}{8} \left(1 + \frac{3}{2}e_1^2\right) (3 cI^2 - 1) +
      \frac{15}{16} e_1^2 sI^2 \cos 2(\Omega_2 - \varpi_1) \label{eq:quad}\\
R_3 &= \frac{1}{64} \left[
\left(-3 + 33 cI + 15cI^2 - 45cI^3\right) b_1 
      \cos (\varpi_2  +\varpi_1 - 2\Omega_2)  \right. \nonumber \\
&+ \left(-3 - 33 cI + 15cI^2 + 45cI^3\right) b_1 
      \cos (\varpi_2  - \varpi_1) \nonumber \\
&+ \left(15 - 15cI -15cI^2 + 15cI^3\right) b_2 
      \cos (\varpi_2  +3 \varpi_1 - 4\Omega_2) \nonumber \\
&+ \left. \left(15 + 15cI -15cI^2 - 15cI^3\right) b_2 
      \cos (\varpi_2  - 3 \varpi_1 + 2\Omega_2) 
\right] \label{eq:oct}\\
R_4 &= \frac{3}{16}  \left( 2 + 3 e_2^2 \right) d_1^\ast - \frac{495}{1024}
  e_2^2 - \frac{135}{256} cI^2 - \frac{165}{512} \nonumber \\
  &+ \frac{315}{512}
  cI^4 + \frac{945}{1024} cI^4 e_2^2 - \frac{405}{512} cI^2e_2^2 
      \nonumber \\
&+ \left( \frac{105}{512} cI^4 + \frac{315}{1024} e_2^2 -
  \frac{105}{256} cI^2 + \frac{105}{512}  - \frac{315}{512}
  cI^2e_2^2 \right .\nonumber \\
  &\left. + \frac{315}{1024} cI^4 e_2^2 \right) d_3 \cos 4(\varpi_1 
  - \Omega_2) \nonumber \\
&+ \frac{105}{512} \left( cI^3 - cI - \frac{1}{2} cI^4 +
  \frac{1}{2} \right) d_3 e_2^2 
\cos (4\varpi_1  +2 \varpi_2 - 6\Omega_2) 
  \nonumber \\
&+ \frac{105}{512} \left( -cI^3 + cI - \frac{1}{2} cI^4 +
  \frac{1}{2} \right) d_3 e_2^2 
\cos (4\varpi_1  -2 \varpi_2 - 2\Omega_2) 
  \nonumber \\
&+ \left( \frac{45}{64}cI^2 - \frac{45}{512} - \frac{315}{512}cI^4 
  \right) e_2^2 
\cos (2\varpi_2 - 2\Omega_2) \nonumber \\
&+ \left( \frac{15}{16} cI^2 + \frac{45}{32} cI^2 e_2^2 -
  \frac{45}{256} e_2^2 - \frac{15}{128} - \frac{315}{256} cI^4 e_2^2 \right.
  \nonumber \\
  &\left. - \frac{105}{128} cI^4 \right) d_2 \cos (2 \varpi_1 - 2 \Omega_2) \nonumber \\
&+ \left( \frac{15}{256} - \frac{45}{128} cI^2 + \frac{75}{256} cI 
  + \frac{105}{256} cI^4 \right. \nonumber \\
  &- \left. \frac{105}{256} cI^3 \right) d_2 e_2^2 
\cos (2\varpi_1 + 2 \varpi_2 - 4\Omega_2)  \nonumber \\
&+ \left( \frac{15}{256} - \frac{45}{128} cI^2 - \frac{75}{256} cI 
  + \frac{105}{256} cI^4 \right. \nonumber \\
  &+ \left. \frac{105}{256} cI^3 \right) d_2 e_2^2 
\cos (2\varpi_1 - 2 \varpi_2)  \label{eq:hex}\\
R &= R_2 c_1 + R_3 c_2 + R_4 c_3 \,.
\end{align}
A few notes: ($i$) this disturbing function
includes only the quadrupole ($R_2c_1$), octopole
($R_3c_2$), and hexadecapole ($R_4c_3$) terms;
the monopole term has been dropped (it is equivalent
to adding $m_1$ to the central mass $m_0$), as has the dipole term 
which orbit-averages to zero;
($ii$) there are typos in equation (6) of Y03:
their $c_i$'s are missing factors of $M_\odot$ ($\rightarrow m_1$);
($iii$) we have starred $d_1^\ast$
in equation (\ref{eq:d1star}) to highlight that this term
as printed in Y03 is in error, as brought to our
attention by Matija \'Cuk, who also provided the correction.
We have verified this correction independently by
computing the hexadecapole disturbing function
in the limit $I=0$ and $e_2 = 0$.

We insert the disturbing function $R$ into Lagrange's planetary
equations for $\dot{e}_2$, $\dot{\Omega}_2$, $\dot{\varpi}_2$, and $\dot{I}$ 
(equations 6.146, 6.148, 6.149, and 6.150 of \citealt{murray00}).
These coupled ordinary differential equations are
solved numerically using a Runge-Kutta-Dormand-Prince method with
stepsize control and dense output ({\tt runge\_kutta\_dopri5} in {\tt C++}).

\subsection{Fixed Parameters} \label{sec:fixed}
The number of parameters is daunting,
even for the restricted, secular three-body problem considered here.
Throughout this paper, we fix the following parameters:
\begin{align}
a_1 &= 20 \, {\rm AU} \\
a_2 &= 100 \, {\rm AU} \\
m_0 &= 1 M_\odot \\
m_1 &= 0.001\, m_0 \,.
\end{align}
The ratio of orbital semimajor axes is fixed at
$\alpha \equiv a_1/a_2 = 0.2$, the largest value we thought
might still be amenable to a truncated expansion in $\alpha$
of the disturbing function (the smaller is $\alpha$,
the better the agreement with $N$-body integrations,
as we have verified explicitly; see also
Section \ref{sec:nbody}). Many of our results---the existences of
secular resonances, and the amplitudes
of eccentricity and inclination variations---fortunately do not depend
critically on $\alpha$; for a different $\alpha$, we
can obtain the same qualitative results by
adjusting initial eccentricities
(see, e.g., Section \ref{sec:circular}).
The above parameter choices do directly determine the timescales
of the test particle's evolution, which should (and mostly do)
fall within the Gyr ages of actual planetary systems
(see Section \ref{sec:prectime}).
Our parameters are those of a
distant Jupiter-mass planet (like 51 Eri b; \citealt{macintosh15})
perturbing an exterior collection of minor bodies (like the Kuiper belt).

With no loss of generality, we align the apsidal line
of the perturber's orbit with the $x$-axis:
\begin{equation}
\varpi_1 = 0^\circ \,.
\end{equation}
The remaining variables of the problem are $e_1$, $e_2$, $I$, $\omega_2$,
and $\Omega_2$. Often in lieu of $e_2$ we will plot the 
periastron distance $q_2 = a_2 (1-e_2)$ to see how close the test particle
comes to the perturber. Once the orbits cross or nearly cross
(i.e., once $q_2 \lesssim a_1 (1+e_1) = 20$--35 AU),
our secular equations break down
and the subsequent evolution cannot be trusted.
Nevertheless we will sometimes show orbit-crossing trajectories
just to demonstrate that channels exist whereby test particle
periastra can be lowered from large distances to
near the perturber (if not conversely).

To the extent that $R$ is dominated by
the first quadrupole term in (\ref{eq:quad})
proportional to $(3 \cos^2 I - 1)$,
more positive $R$ corresponds to more co-planar orbits (i.e.,
wires 1 and 2 closer together).
The numerical values for $R$ quoted below
have been scaled to avoid large unwieldy numbers;
they should be multiplied by $3.55 \times 10^6$
to bring them into cgs units.

\section{Circular Perturber} \label{sec:circular}
When $e_1=0$, the octopole contribution to the potential vanishes, but the
quadrupole and hexadecapole contributions do not.

Because the potential
for a circular perturber is axisymmetric, the $z$-component of the test
particle's angular momentum is conserved (we omit the subscript 2 on
$J_z$ for convenience):
\begin{equation} \label{eq:jz}
J_{z} \equiv \sqrt{1-e_2^2} \cos I
\end{equation}
where we have dropped the dependence of the angular momentum on $a_2$,
since semimajor axes never change in secular dynamics.
We therefore have two constants of the motion: $J_{z}$ and the disturbing
function itself (read: Hamiltonian), $R$.
Smaller $J_{z}$ corresponds to more highly inclined and/or more eccentric
test particle orbits.

\subsection{The Inverse Kozai ($\omega_2$) Resonance} \label{sec:ikr}
\begin{figure*}
	\includegraphics[width=\textwidth]{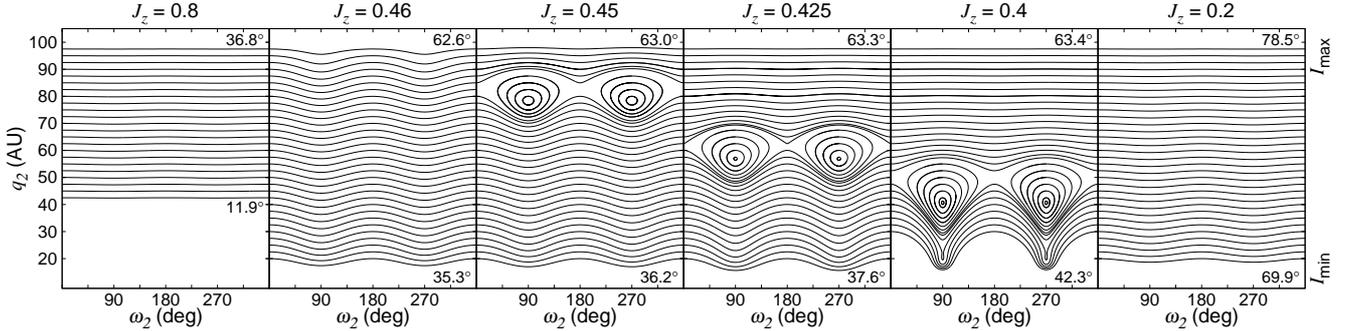}
    \caption{Periastron distances $q_2$ vs.~arguments of periastron $\omega_2$, for $e_1 = 0$. Because the potential presented by a circular perturber is axisymmetric,
     $J_z = \sqrt{1-e_2^2} \cos I$ is conserved; trajectories in a given panel
    have $J_z$ as annotated. For given $J_z$, the
    inclination $I$ increases monotonically but not linearly
    with $q_2$; the maximum and minimum inclinations
    for the trajectories plotted are labeled on the right of each panel.
    In a narrow range of $J_z = 0.40$--0.45 (for our chosen
    $\alpha = a_1/a_2 = 0.2$), the inverse Kozai (a.k.a.~$\omega_2$)
    resonance appears, near $I \simeq 63^\circ$.
    Near $J_z = 0.40$, the inverse Kozai resonance can force the test particle to cross orbits with the perturber.}
    \label{fig1}
\end{figure*}

Figure \ref{fig1} gives a quick survey of the test particle dynamics
for $e_1 = 0$.
For a restricted range in $J_z \simeq $ 0.40--0.45,
the test particle's argument of periastron $\omega_2$
librates about either $90^\circ$ or 270$^\circ$,
with concomitant oscillations
in $q_2$ (equivalently $e_2$) and $I$. This is the
analogue of the conventional Kozai resonance, exhibited
here by an exterior test particle;
we refer to it as the ``inverse Kozai'' resonance
or the $\omega_2$ resonance. The inverse Kozai resonance
appears only at hexadecapole order; it originates from the term
in (\ref{eq:hex}) proportional to
$e_2^2 \cos (2\varpi_2 - 2\Omega_2) = e_2^2 \cos 2\omega_2$.\footnote{\citet{naoz17} refer to their octopole-level treatment as exploring the ``eccentric Kozai-Lidov mechanism for an outer test particle.'' Our terminology here differs; we consider the analogue of the Kozai-Lidov resonance the $\omega_2$ resonance, which appears only at hexadecapole order, not the $\Omega_2-\varpi_1$
resonance that they highlight.}

The inverse Kozai resonance appears near 
\begin{equation}
I (\omega_2{\rm-res})
= \arccos (\pm \sqrt{1/5}) \simeq 63^\circ \, {\rm and} \, 117^\circ
\end{equation}
which, by Lagrange's planetary equations and (\ref{eq:quad}),
are the special inclinations
at which the quadrupole precession rate
\begin{equation}
\left. \frac{d\omega_2}{dt} \right|_{{\rm quad},e_1=0} = \frac{3}{8} \frac{m_1}{m_0} \left( \frac{a_1}{a_2} \right)^2 \frac{n_2}{(1-e_2)^2} \left( 5 \cos^2 I - 1 \right) \label{eq:domegadt}
\end{equation}
vanishes, where $n_2$ is the test particle mean motion;
see Gallardo et al.~(\citeyear{gallardo12}, their equation 11 and
subsequent discussion). At $I = I (\omega_2{\rm-res})$, fixed points
appear at $\omega_2 = 90^\circ$ and $\omega_2 = 270^\circ$.
The critical angles $63^\circ$ and $117^\circ$ are related
to their well-known Kozai counterparts 
of $39^\circ$ and $141^\circ$ (i.e., $\arccos (\pm \sqrt{3/5})$),
but the correspondence is not exact. In the conventional
problem, the inclinations at which the fixed points
($\omega_1 = \pm 90^\circ$, $\dot{\omega}_1=0$) appear vary
from case to case; they are given by
$I = \arccos \left[ \pm \sqrt{(3/5) (1-e_1^2)} \right] = \arccos \left[ \pm (3/5)^{1/4} |J_{1z}|^{1/2} \right]$,
where $J_{1z} \equiv \sqrt{1-e_1^2} \cos I$ is conserved
at quadrupole order (e.g., \citealt{lithwick11}).
But for our inverse problem, the fixed points ($\omega_2 = \pm 90^\circ$,
$\dot{\omega}_2 = 0$) appear at fixed inclinations $I$ of $63^\circ$
and $117^\circ$ that are independent of $J_z$ (for $e_1 = 0$).
In this sense, the inverse Kozai resonance is ``less flexible''
than the conventional Kozai resonance.

The $\omega_2$ resonance exists only in a narrow range
of $J_z$ that is
specific to a given $\alpha = a_1/a_2$, as we have determined
by numerical experimentation. 
Outside this range,
$\omega_2$ circulates and $e_2$ and $I$ hardly vary (Figure \ref{fig1}).
Fine-tuning $J_z$ can produce large
resonant oscillation amplitudes in $e_2$ and $I$;
some of these trajectories lead to orbit crossing
with the perturber, as seen in the panel for $J_z = 0.400$
in Figure \ref{fig1}.

\begin{figure}
	\includegraphics[width=\columnwidth]{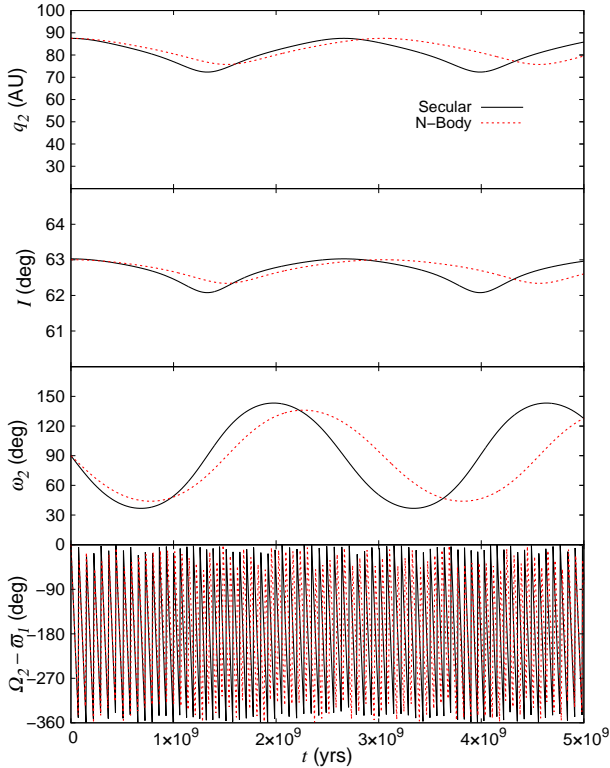}
    \caption{Time evolution within the inverse Kozai resonance.
    The trajectory chosen is the one in Figure \ref{fig1} with
    $J_z = 0.45$ and the largest libration amplitude.
    The nodal ($\Omega_2$) precession arises from the
    quadrupole potential and is therefore on the order of
    $1/\alpha^2 \sim 25$ times faster than the libration
    timescale for $\omega_2$, which is determined
    by the hexadecapole potential.
    Initial conditions: 
    $\varpi_2 = 90^\circ$,
    $\Omega_2 = 0^\circ$, 
    $q_2 = 87.5$ AU, $I = 63^\circ$. 
Overplotted in red dashed lines are the results of an $N$-body integration 
with identical initial conditions (and initial true anomalies
$f_1 = 0^\circ$ and $f_2 = 180^\circ$) inputted as Jacobi coordinates.
The $N$-body integration is carried out
using \texttt{WHFast}, part of the \texttt{REBOUND} package
(\citealt{rein15}; \citealt{wisdom91}).}
    \label{fig:gallardo_time}
\end{figure}

\subsubsection{Precession Timescales} \label{sec:prectime}
To supplement (\ref{eq:domegadt}),
we list here for ease of reference the remaining equations of motion
of the test particle, all to leading order,
as derived by \citet{gallardo12}
for the case $e_1 = 0$.
We have verified that the disturbing function
we have derived in Section \ref{sec:disturb} yields identical expressions:
\begin{align}
\left. \frac{de_2}{dt} \right|_{{\rm hex}, e_1=0} &= +\frac{45}{512} \frac{m_1}{m_0} \left( \frac{a_1}{a_2} \right)^4 \frac{ e_2 n_2 }{(1-e_2^2)^3} \nonumber \\
& \times \left( 5 + 7 \cos 2I \right) \sin^2 I \sin 2\omega_2\\
\left. \frac{dI}{dt} \right|_{{\rm hex}, e_1 = 0} &= -\frac{45}{1024} \frac{m_1}{m_0} \left( \frac{a_1}{a_2} \right)^4 \frac{ e_2^2 n_2 }{(1-e_2^2)^4} \nonumber \\
& \times \left( 5 + 7 \cos 2I \right) \sin 2I \sin 2\omega_2 \\
\left. \frac{d\Omega_2}{dt} \right|_{{\rm quad}, e_1 = 0} &= -\frac{3}{4} \frac{m_1}{m_0} \left( \frac{a_1}{a_2} \right)^2 \frac{ n_2 }{(1-e_2^2)^2} \cos I \label{eq:dOmegadt}\\
\left. \frac{d\varpi_2}{dt} \right|_{{\rm quad}, e_1 = 0} &= +\frac{3}{16} \frac{m_1}{m_0} \left( \frac{a_1}{a_2} \right)^2 \frac{ n_2 }{(1-e_2^2)^2} \left( 3 - 4 \cos I + 5 \cos 2I \right) \,. \label{eq:dvarpidt}
\end{align}
As equations (\ref{eq:dOmegadt}) and (\ref{eq:dvarpidt}) show,
the magnitudes of the precession rates
for $\Omega_2$ and $\varpi_2$ are typically similar
to within order-unity factors. We define a fiducial secular
precession period 
\begin{equation} \label{eq:fiducial}
\left. t_{\rm prec} \right|_{{\rm quad}, e_1=0} \sim \frac{2\pi}{n_2} \frac{m_0}{m_1} \left( \frac{a_2}{a_1} \right)^2 \left( 1-e_2^2 \right)^2
\end{equation}
which reproduces the precession period for $\Omega_2$
seen in the sample evolution
of Figure \ref{fig:gallardo_time} to within a factor of 3.
The scaling factors in (\ref{eq:fiducial}) are more reliable than the
overall magnitude; the dependencies on $m_0$, $m_1$, $a_1$, and $a_2$
can be used to scale the time coordinate
of one numerical computation to another.
Figure \ref{fig:gallardo_time} is made for a particle in the inverse
Kozai resonance; note how the oscillation periods for $\omega_2$, and
by extension $I$ and $q_2$, are each a few dozen times longer
than the nodal precession period. This is expected since for the
inverse Kozai resonance, $d\omega_2/dt$ vanishes at quadrupole order,
leaving the hexadecapole contribution, which is smaller by
$\sim$$(a_1/a_2)^2 = 1/25$, dominant.

As shown in Figure \ref{fig:gallardo_time},
the secular trajectory within the $\omega_2$ resonance
is confirmed qualitatively by the $N$-body 
symplectic integrator \texttt{WHFast} (\citealt{rein15}; \citealt{wisdom91}), part of the \texttt{REBOUND} package (version 3.5.8; \citealt{rein12}). A timestep of 0.25 yr was used (0.28\% of the orbital
period of the interior perturber) for the $N$-body integration shown;
it took less than 3 wall-clock hours to complete the 5 Gyr integration using
a 2.2 GHz Intel Core i7 processor on a 2015 MacBook Air laptop.

\subsubsection{Inverse Kozai vs.~Kozai} \label{sec:ikk}
\begin{figure}
	\includegraphics[width=\columnwidth]{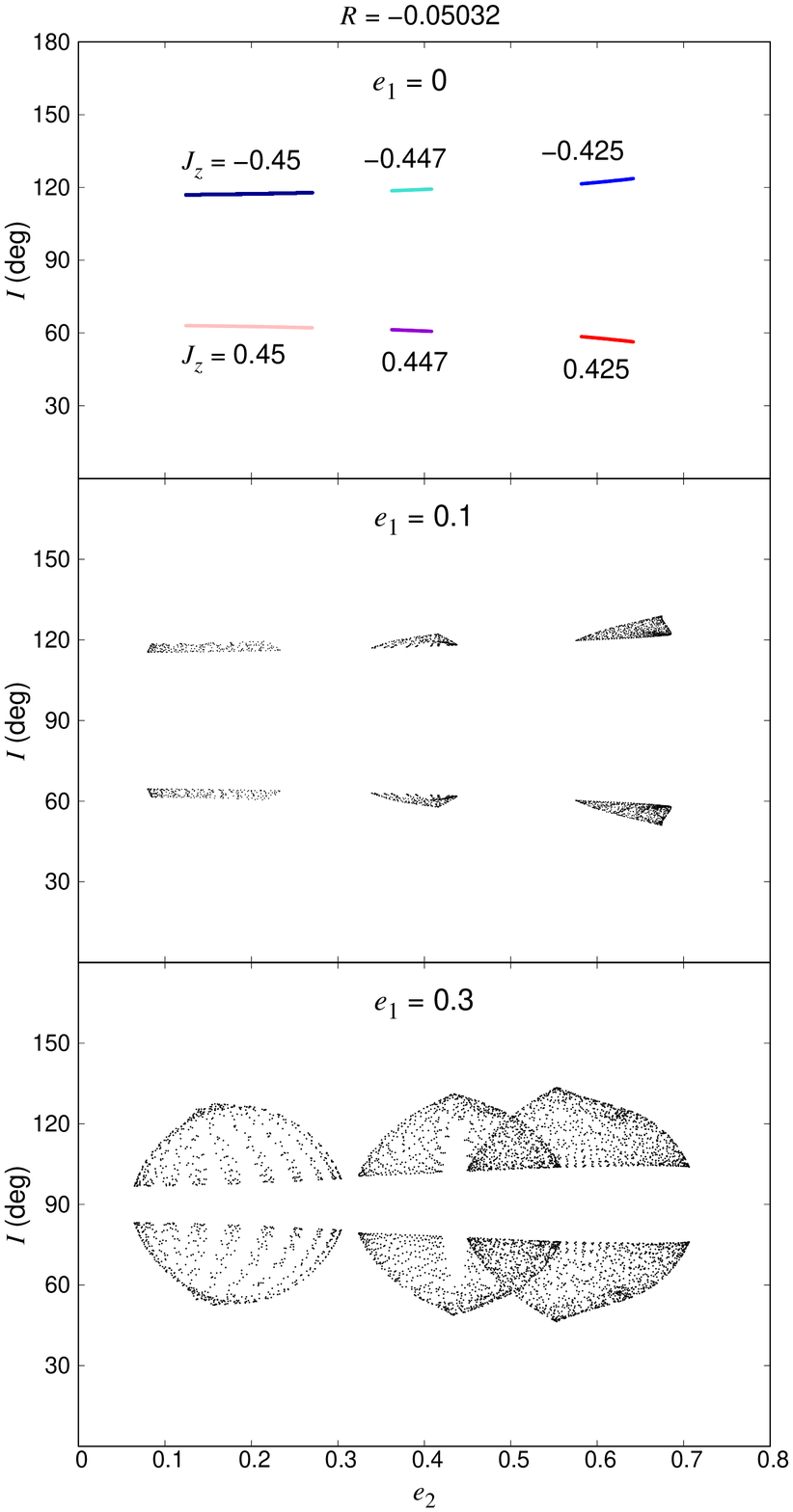}
    \caption{
    Inclination $I$ vs.~eccentricity $e_2$ for a constant disturbing function
    $R = -0.05032$ (see Section \ref{sec:fixed} for the units of $R$).
   When $e_1 = 0$, $J_z$ is an additional constant of the motion;
   the resultant ``inverse Kozai curves'' (top panel) for our external test
   particle are analogous to the conventional ``Kozai curves'' shown in
   Figure 2 of \citet{lithwick11} for an internal test particle.
   Compared to the conventional case, the ranges of $I$ and $e_2$ in the
   inverse case are much more restricted; what variation there is
   is only possible because of the $e_2^2 \cos (2\omega_2)$ term
   that appears at hexadecapolar order (see equation \ref{eq:hex}).
   As $e_1$ increases above zero (middle and bottom panels), 
   $J_z$ varies more and variations in $I$ grow larger.
   In each of the middle and bottom panels, points are generated
   by integrating the equations of motion for six sets of initial
   conditions specified in Table \ref{tab:inverse_Kozai_curves}.}
    \label{fig:inverse_Kozai_curves}
\end{figure}

In the top panel of Figure \ref{fig:inverse_Kozai_curves}, we show
analogues to the ``Kozai curves'' 
made by \citet{lithwick11} for the conventional problem.
This top panel delineates the allowed values of test particle
eccentricity and inclination for given $J_z$ and $R$ when $e_1 = 0$.
Contrast these ``inverse Kozai curves'' with the Kozai
curves calculated by \citet{lithwick11} in their Figure 2 (left panel):
for the inverse problem, the range of allowed
eccentricities and inclinations is much
more restricted (at fixed $J_z$ and $R$) 
than for the conventional problem.
For the inverse problem when $e_1 = 0$,
$e_2$ and $I$ are strictly constant at quadrupole order;
variations in $e_2$ and $I$ for the case of a circular perturber 
are possible starting only at hexadecapole order,
via the small inverse Kozai resonant term
in $R$ proportional to $e_2^2 \cos 2\omega_2$ (variations in $\omega_2$
directly drive the variations in $e_2$ and $I$ 
when $e_1 = 0$).
By comparison, in the conventional problem, variations in test particle
eccentricity and inclination are possible even at quadrupole order, and large.

This key difference between the conventional and inverse
problems stems from the difference between the interior
and exterior expansions of the $1/\Delta$ potential.
The conventional interior expansion involves the sum
$P_\ell r_{\rm test}^{\ell}$, 
where $P_\ell$ is the Legendre polynomial of order $\ell$ 
and $r_{\rm test}$ is the radial position of the test particle.
The inverse exterior expansion
involves a qualitatively different sum,
$P_\ell r_{\rm test}^{-(\ell+1)}$.
The time-averaged potentials in
the conventional and inverse problems therefore involve
different integrals; 
what averages to a term proportional to
$\cos(2\omega_{\rm  test})$
in the conventional quadrupole problem averages 
instead in the inverse problem to a constant,
independent of the test particle's argument of periastron
$\omega_{\rm test}$ (we have verified this last statement by evaluating
these integrals).
An interior multipole moment of order $\ell$
is not the same as an exterior multipole moment of the same order.
We could say that the inverse exterior potential looks ``more Keplerian''
insofar as its monopole term scales as $1/r_{\rm test}$.

\section{Eccentric Perturber} \label{sec:eccentric}

When $e_1 \neq 0$, all orders (quadrupole, octopole, and hexadecapole)
contribute to the potential seen by the test particle.
The potential is no longer axisymmetric, and so $J_z$ is no longer
conserved. This opens the door to orbit ``flipping'', i.e.,
a prograde ($I < 90^\circ$) orbit can switch to being retrograde
($I > 90^\circ$) and vice versa (e.g., \citealt{naoz17}).
There is only one constant of the motion, $R$.

\subsection{First Survey}
Whereas when $e_1=0$ the evolution did not depend on
$\Omega_2$, it does when $e_1 \neq 0$. For our first foray
into this large multi-dimensional phase space, we divided up
initial conditions as follows. For
each of four values of $e_1 \in \{0.03, 0.1, 0.3, 0.7\}$,
we scanned systematically through
different initial values of $q_{2,{\rm init}}$
(equivalently $e_{2,{\rm init}}$) 
ranging between $a_2 = 100$ AU and $a_1 = 20$ AU.
For each $q_{2,{\rm init}}$, we assigned $I_{\rm init}$ according
to one of three values of $J_{z,{\rm init}} \equiv \sqrt{1-e_{2,{\rm init}}^2} \cos I_{\rm init} \in \{0.8, 0.45, 0.2\}$,
representing ``low'', ``intermediate'', and ``high'' inclination
cases, broadly speaking.
Having set $e_1$, $q_{2,{\rm init}} (e_{2,{\rm init}})$,
and $J_{z,{\rm init}} (I_{\rm init})$, we cycled through five values of
$\varpi_{2,{\rm init}} \in \{0^\circ, 45^\circ, 90^\circ, 135^\circ, 180^\circ\}$
and three values of $\omega_{2,{\rm init}} \in \{0^\circ,90^\circ,270^\circ\}$.

We studied all integrations from this large ensemble, adding more 
with slightly different initial conditions as our curiosity led us.
In what follows, we present a subset of the results from this first
survey, selecting those we thought representative or interesting.
Later, in Section \ref{sec:sos}, we will provide a second and
more thorough survey using surfaces of section.
A few sample integrations from both surveys will be tested
against $N$-body calculations in Section \ref{sec:nbody} (see also
Figure \ref{fig:gallardo_time}).

\subsubsection{Low Perturber Eccentricity $e_1 \leq 0.1$}
\begin{figure}
	\includegraphics[width=\columnwidth]{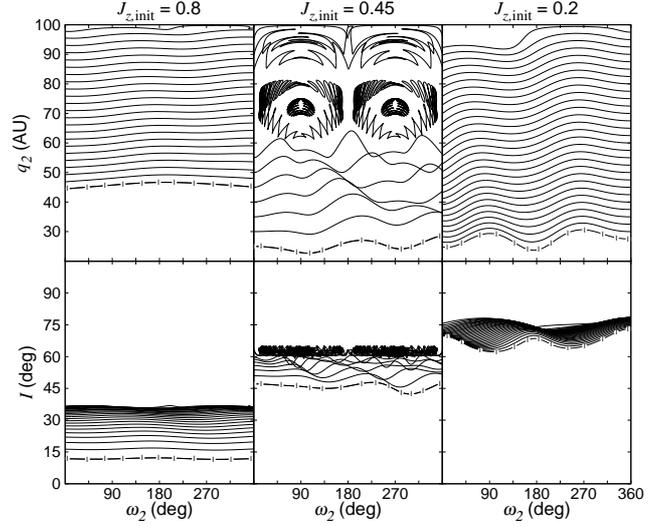}
    \caption{Analogous to Figure \ref{fig1},
    but now for a mildly eccentric perturber ($e_1 = 0.1$). 
    Because $e_1 \neq 0$, $J_z$ is not conserved and cannot be used
    to connect $q_2$ and $I$ uniquely; we have to plot
    $q_2$ and $I$ in separate panels.
    Nevertheless, $e_1$ is still small enough that $J_z$ is approximately
    conserved; $q_2$ and $I$ still roughly follow one another
    for a given $J_{z,{\rm init}}$, i.e., the family of trajectories
    proceeding from lowest $I$ (marked by vertical bars) to highest $I$
    corresponds to the same family of trajectories proceeding from lowest
    $q_2$ (marked by vertical bars) to highest $q_2$.
    The $\omega_2$ resonance can still
    be seen near $I \simeq 63^\circ$,
    in the center panels for $J_{z,{\rm init}} = 0.45$.
    All the non-resonant trajectories are
    initialized with $\varpi_2 = 0^\circ$ and $\omega_2 = 0^\circ$.
    For the four resonant
    trajectories, the initial $\varpi_2 = 0^\circ$ and
    $\omega_2 = \pm 90^\circ$.
    }
    \label{fig:different_jz}
\end{figure}

Comparison of Figure \ref{fig:different_jz} with Figure \ref{fig1} shows that at low
perturber eccentricity, $e_1 \lesssim 0.1$, the test particle
does not much change its behavior from when $e_1 = 0$
(for a counter-example, see Figure 6).
The same inverse Kozai resonance appears for $J_{z,{\rm init}} = 0.45$
and $e_1 = 0.1$ as it does for $e_1 = 0$. The maximum libration
amplitude of the resonance is somewhat higher at the larger $e_1$.
The trajectories
shown in Figure \ref{fig:different_jz} are for
$\varpi_{2,{\rm init}} = 0^\circ$,
but qualitatively similar results obtain for other choices
of $\varpi_{2,{\rm init}}$.

The middle panel of Figure \ref{fig:inverse_Kozai_curves}
elaborates on this result,
showing that even though $J_{z,{\rm init}}$
is not strictly conserved when $e_1\neq 0$, it can be approximately
conserved (again, see the later Figure \ref{fig:flip}
for a counter-example).
Test particles explore
more of $e_2$-$I$ space when $e_1 = 0.1$ than when $e_1 = 0$,
but they still largely
respect (for the specific $R$ of Figure \ref{fig:inverse_Kozai_curves})
the constraints imposed by $J_z$ when $e_1 = 0$.
This statement also holds at $e_1 = 0.3$ (lower panel), but to a lesser
extent.

\begin{figure}
	\includegraphics[width=\columnwidth]{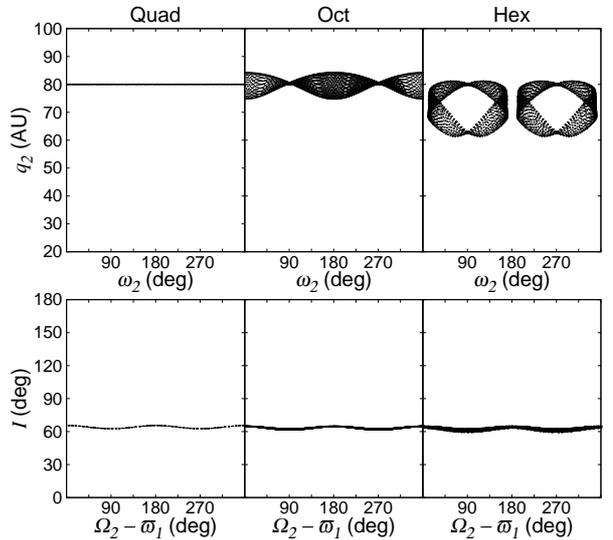}
    \caption{
    Comparing quadrupole (quad), octopole (oct), and
    hexadecapole (hex) evolutions for $e_1 = 0.1$ and $\varpi_1 = 0^\circ$
    and the same test particle initial 
    conditions ($e_2 = 0.2$, $I = 62.66^\circ$,
    $\varpi_2 = 0^\circ$, and $\omega_2 = 90^\circ$).
    The hex panel features a second set of initial conditions
    identical to the first except that $\omega_2 = 270^\circ$;
    the two hex trajectories map to the same quad and oct trajectories as shown.
    The inverse Kozai resonance, featuring libration of
    $\omega_2$ about $\pm 90^\circ$ and $\sim$20\%
    variations in $e_2$, appears only in a 
    hex-level treatment.
    }
    \label{fig:gallardo_hex_v_oct}
\end{figure}

Figure \ref{fig:gallardo_hex_v_oct}
illustrates how the hexadecapole (hex) potential---specifically
the inverse Kozai resonance---can qualitatively change
the test particle dynamics at octopole (oct) order.
Only at hex order is the $\omega_2$ resonance evident.
Compared to the oct level dynamics, the periastron distance
$q_2$ varies more strongly, hitting its maximum and minimum values
at $\omega_2 = 90^\circ$ or $270^\circ$
(instead of at $0^\circ$ and $180^\circ$,
as an oct treatment would imply).

Orbit flipping becomes possible when $e_1 \neq 0$, for sufficiently
large $I$ or $e_2$ \citep{naoz17}. Figure \ref{fig:flip} is analogous to Figure \ref{fig:inverse_Kozai_curves}
except that it is made for a more negative $R$, corresponding 
to larger $I$ (insofar as $R$ is dominated by the quadrupole
term). For this $R = -0.1373$, as with the previous $R = -0.05032$,
$e_2$ and $I$
hardly vary when $e_1 = 0$ (Section \ref{sec:ikk}).
But when $e_1 = 0.1$, the constraints imposed by fixed $J_z$
come loose; Figure \ref{fig:flip} shows that a single particle's $J_z$
can vary dramatically from positive (prograde) to negative (retrograde)
values. As shown by Naoz et al.~(\citeyear{naoz17}, see their Figure 1),
such orbit flipping
is possible even at quadrupole order; flipping is not associated
with the $\omega_2$ resonance, but rather with librations of 
$\Omega_2-\varpi_1$ 
about $90^\circ$ or $270^\circ$.
We verify the influence of this
$\Omega_2-\varpi_1$ resonance
in the middle panel of Figure \ref{fig:flip}.

\begin{figure}
	\includegraphics[width=\columnwidth]{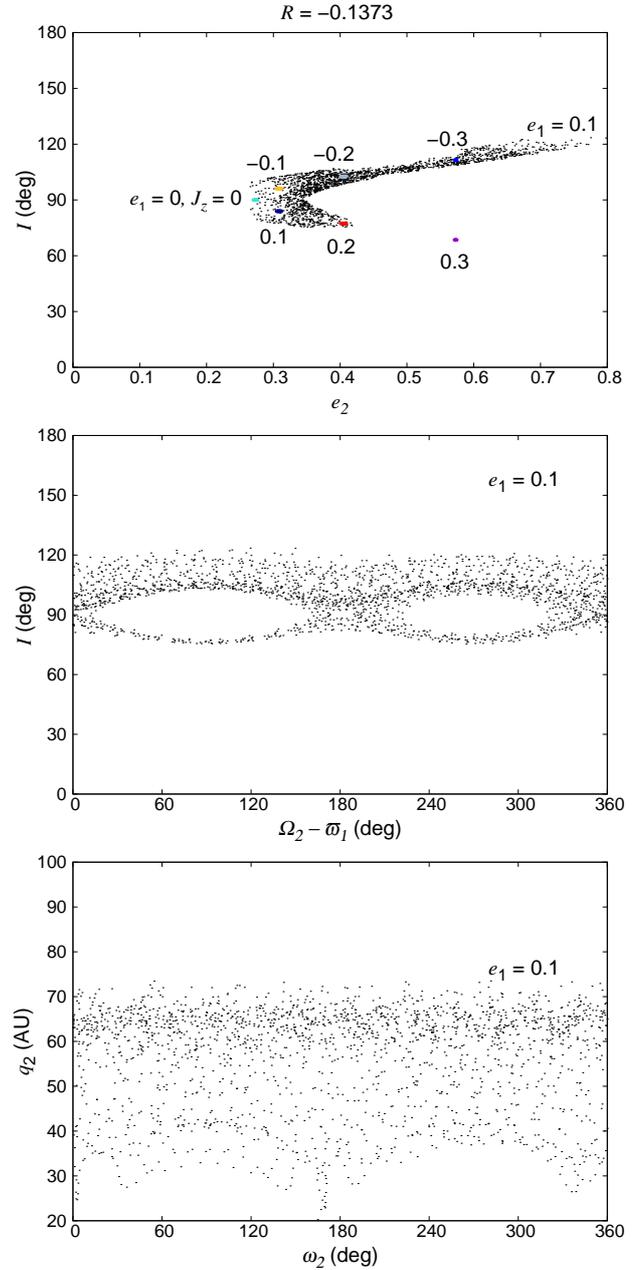}
    \caption{Top panel: Inclination $I$ vs.~eccentricity $e_2$ for a fixed disturbing function $R = -0.1373$ (see Section \ref{sec:fixed} for the units of $R$). 
    Different colored points, corresponding
    to different $J_z$ values as marked, 
    are for $e_1 = 0$, and are analogous to those shown in the top panel of
    Figure \ref{fig:inverse_Kozai_curves}. The black points represent the
    trajectory of a single test particle, integrated for $e_1 = 0.1$ 
    and using the following initial conditions:
    $e_2 = 0.3691$, $I = 85^\circ$, $\varpi_2 = 0^\circ$,
    and $\Omega_2 = 0^{\circ}$.
    When $e_1 \neq 0$, $J_z$ is no longer conserved, and $e_2$ and $I$
    vary dramatically for this value of $R$; $J_z$ even changes sign
    as the orbit flips.
    The variation in $e_2$ is so large that eventually
    the test particle crosses the orbit of the perturber ($e_2 > 0.8$),
    at which point we terminate the trajectory.
    Center panel: Inclination $I$ vs.~longitude of ascending node $\Omega_2$
    (referenced to $\varpi_1$, the periapse longitude of the perturber)
    for the
    same
    black trajectory shown in the top panel.
    The two lobes of the $\Omega_2-\varpi_1$ resonance \citep{naoz17},
    around which the
    particle lingers, are visible.
    Bottom panel: The same test particle trajectory shown in black
    for the top and middle panels, now in $q_2$ vs.~$\omega_2$ space.
    The evolution is evidently chaotic.
    }
    \label{fig:flip}
\end{figure}

\subsubsection{High Perturber Eccentricity $e_1 = 0.3, 0.7$}
\begin{figure}
	\includegraphics[width=\columnwidth]{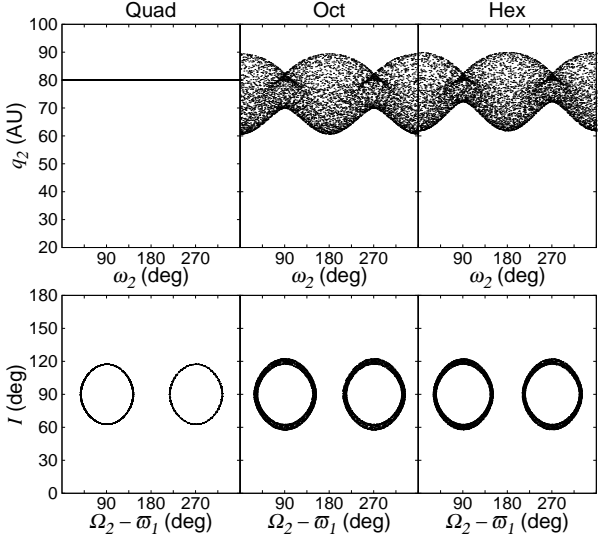}
    \caption{Analogous to Figure \ref{fig:gallardo_hex_v_oct}, but for
    $e_1 = 0.3$ and the following test particle initial
    conditions: $e_2 = 0.2$, 
    $I = 62.66^\circ$,
    $\Omega_2 = \pm 90^\circ$,
    $\varpi_2 = 0^\circ$.
    The two test particle
    trajectories
    overlap in $q_2$-$\omega_2$ space (top panels).
    For these initial conditions,
    the $\Omega_2-\varpi_1$ resonance \citep{naoz17} appears at all orders
    quad through hex (bottom panels). 
    The oct and hex trajectories appear qualitatively
    similar in all respects.
    }
    \label{fig:no_diff}
\end{figure}

We highlight a few comparisons between an oct level
treatment and a hex level treatment. We begin with Figure \ref{fig:no_diff}
which shows practically no difference. Many of the integrations
in our first survey showed no significant difference in going
from oct to hex. We also tested some of the cases
showcased in \citet{naoz17} and found that including the hex
dynamics did not substantively alter their evolution. 

\begin{figure}
	\includegraphics[width=\columnwidth]{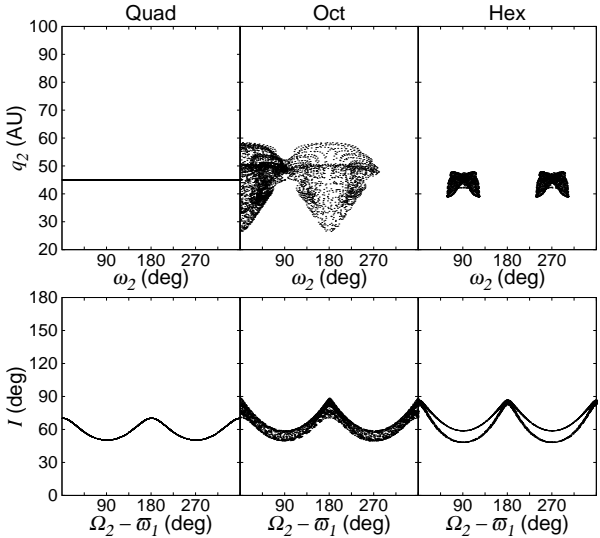}
    \caption{Analogous to Figure \ref{fig:gallardo_hex_v_oct}, but for
    $e_1 = 0.3$ and $\varpi_1 = 0^\circ$
    and the following test particle initial conditions:
    $e_2 = 0.55$, $I = 57.397^\circ$,
    $\Omega_{2} = 45^\circ$ and $225^\circ$,
    $\varpi_2 = 135^\circ$. The inverse Kozai ($\omega_2$)  resonance
    is visible in the hex panels only, with a
    more widely varying inclination here for $e_1 = 0.3$ than
    for $e_1 = 0$ (compare with Figure
    \ref{fig:gallardo_hex_v_oct}).
    The phase space available to the $\omega_2$ resonance shrinks
    with increasing $e_1$; at $e_1 = 0.7$, we could not find
    the resonance (see Figure \ref{fig:SOS_e7}). 
    Two test particle trajectories are displayed for the hex
    panel; since they overlap at the quad and oct levels,
    only one trajectory is shown for those panels
    (the one for which the initial $\Omega_2 = 45^\circ$).}
    \label{fig:gallardo_stab}
\end{figure}

\begin{figure}
	\includegraphics[width=\columnwidth]{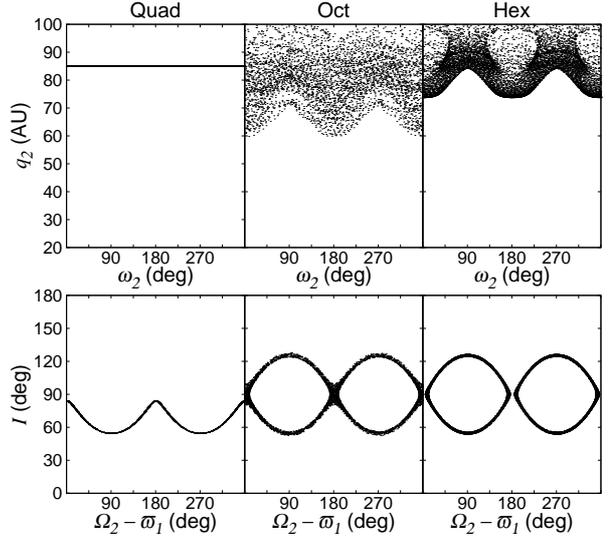}
    \caption{Analogous to Figure \ref{fig:gallardo_hex_v_oct}, but for
    $e_1 = 0.3$ and $\varpi_1 = 0^\circ$
    and the following test particle initial conditions:
    $e_2 = 0.15$, $I = 62.925^\circ$, $\Omega_{2} = 45^\circ$ and
    $225^\circ$, $\varpi_2 = 135^\circ$. 
    Two test particle trajectories are displayed for the hex
    panel; since they overlap at the quad and oct levels,
    only one trajectory is shown for those panels
    (the one for which the initial $\Omega_2 = 225^\circ$).
    The hex potential suppresses the eccentricity variation
    seen at the oct level, and removes the particle
    from the separatrix of the $\Omega_2$ resonance,
    bringing it onto one of two islands of libration.}
    \label{fig:hex_damp1}
\end{figure}

\begin{figure}
	\includegraphics[width=\columnwidth]{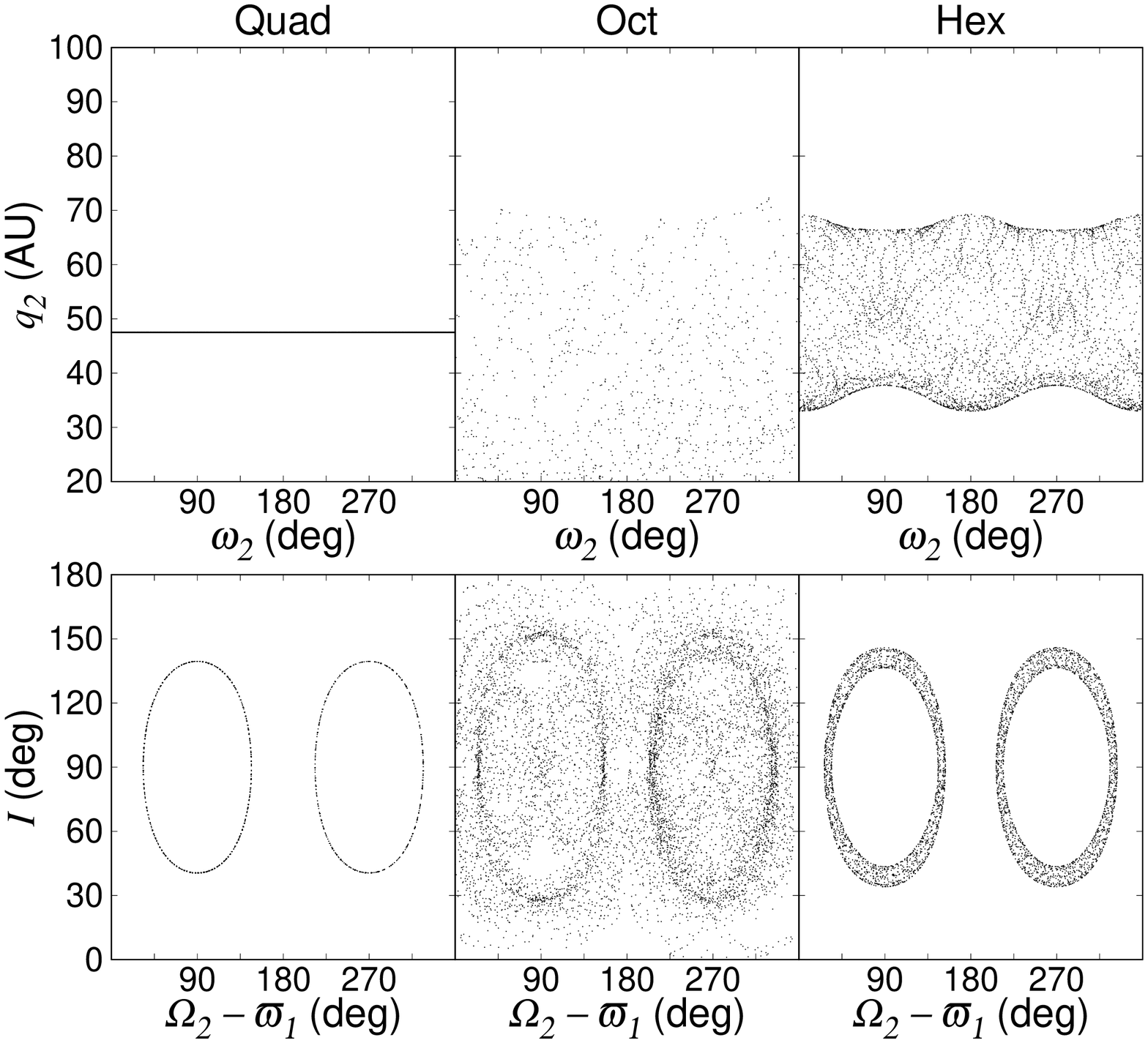}
    \caption{Analogous to Figure \ref{fig:gallardo_hex_v_oct}, but for
    $e_1 = 0.7$ and $\varpi_1 = 0^\circ$
    and the following test particle initial conditions:
    $e_2 = 0.525$, $I = 58.081^\circ$, $\Omega_{2} = 135^\circ$ and
    $315^\circ$, $\varpi_2 = 135^\circ$.
    As with Figures \ref{fig:gallardo_stab} and \ref{fig:hex_damp1},
    the hex potential helps to stabilize the motion; here it
    locks the particle to one of two librating islands of the
    $\Omega_2$ resonance and prevents the orbit crossing
    seen at the oct level.}
    \label{fig:hex_damp2}
\end{figure}

\begin{figure}
	\includegraphics[width=\columnwidth]{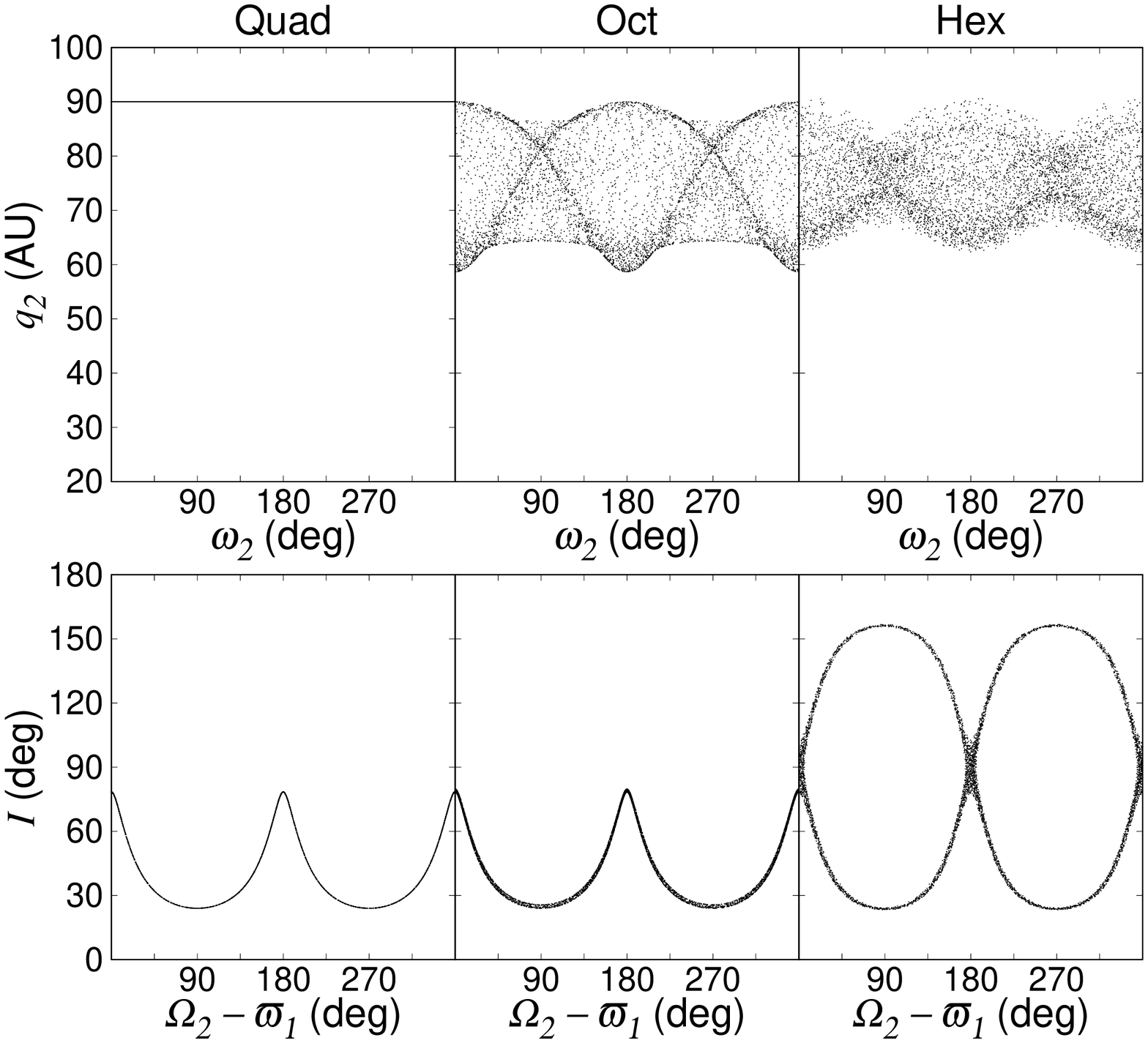}
    \caption{Analogous to Figure \ref{fig:gallardo_hex_v_oct}, but for
    $e_1 = 0.7$ and $\varpi_1 = 0^\circ$
    and the following test particle initial conditions:
    $e_2 = 0.1$, $I = 84.232^\circ$, $\Omega_{2} = 180^\circ$, 
    $\varpi_2 = 180^\circ$. Here the hex potential nudges the particle
    from a circulating trajectory onto the separatrix of the $\Omega_2-\varpi_1$
    resonance (contrast with Figures \ref{fig:hex_damp1} and
    \ref{fig:hex_damp2}).}
    \label{fig:nudge_retro}
\end{figure}

Cases where the hex terms matter are shown in Figures \ref{fig:gallardo_stab}--\ref{fig:nudge_retro}.
The $\omega_2$ resonance, seen only at hex order, can stabilize
the motion; in Figure \ref{fig:gallardo_stab}, the $\omega_2$ resonance eliminates
the chaotic variations seen at the oct level in $q_2$ and $I$.
Even when the $\omega_2$ resonance is not active,
hex level terms can dampen eccentricity and inclination
variations (Figures \ref{fig:hex_damp1} and \ref{fig:hex_damp2}). But the hex terms do not necessarily
suppress; in Figure \ref{fig:nudge_retro} they are seen to nudge the test particle
from a prograde to a retrograde orbit,
across the separatrix of the $\Omega_2-\varpi_1$ resonance.

\subsection{Second Survey: Surfaces of Section} \label{sec:sos}
Surfaces of section (SOS's) afford a more global (if also more
abstract) view of the dynamics. By plotting the test particle's
position in phase space only when one of its coordinates
periodically equals some value, we thin its trajectory out,
enabling it to be compared more easily with
the trajectories of other test particles with different initial
conditions. In this lower dimensional projection, it is also possible
to identify resonances, and to distinguish chaotic from regular
trajectories.

\begin{figure*}
	\includegraphics[width=\textwidth]{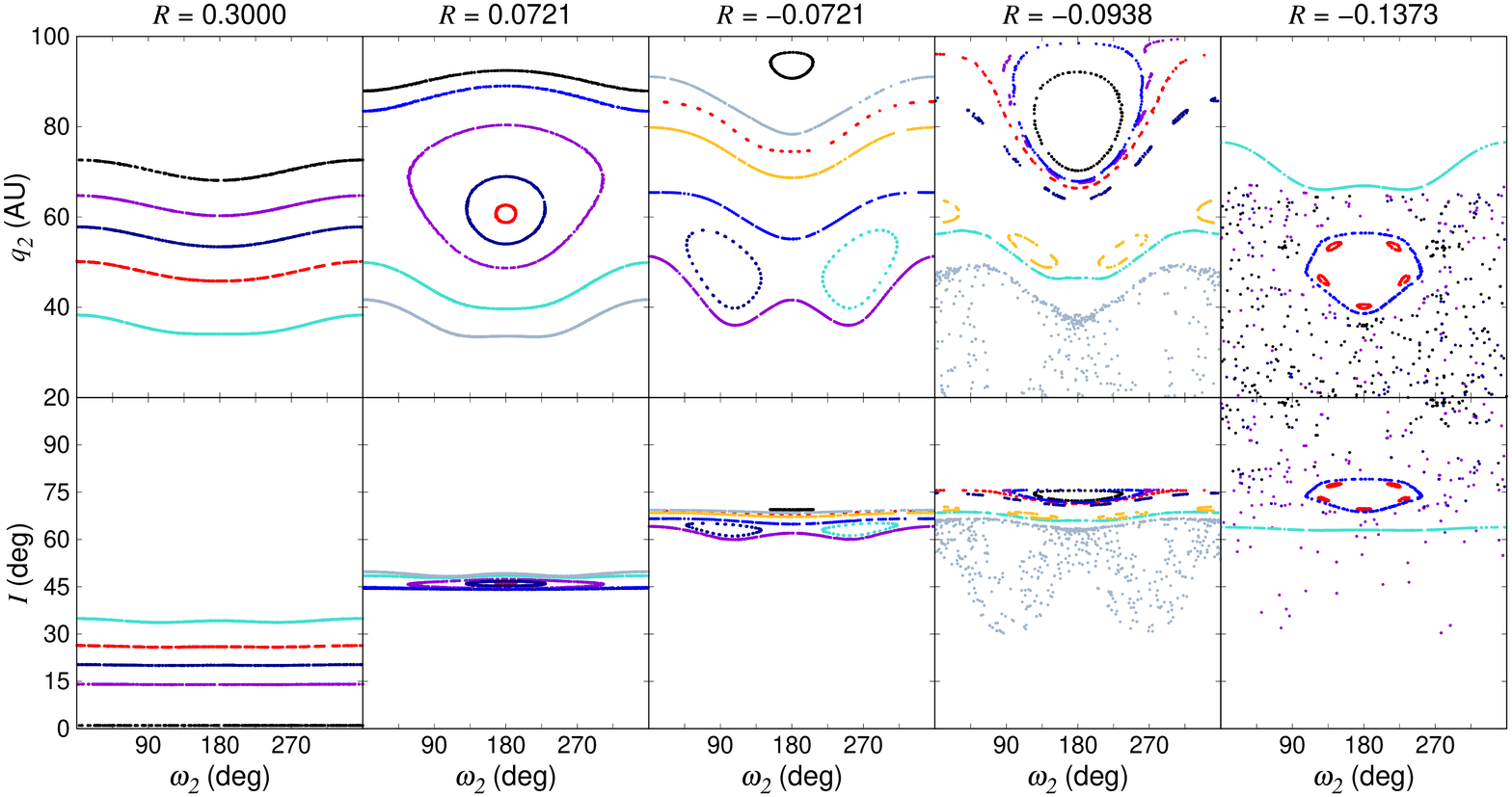}
    \caption{Surfaces of section (SOS's)
    for perturber eccentricity $e_1 = 0.1$ and $\varpi_1 = 0^\circ$ 
    and various values
    of the disturbing function $R$ (the only constant of the motion
    when $e_1\neq 0$; see Section \ref{sec:fixed} for the units of $R$)
    labeled at the top of the figure.
    These SOS's are sectioned using $\Omega_2$: a point is plotted
    every time $\Omega_2$ crosses 180$^\circ$, irrespective of
    the sign of $\dot{\Omega}_2$ (see text). 
    Each test particle trajectory is assigned its own color; see 
    Table \ref{tab:SOS_e1} in the Appendix for the initial conditions.
    At $R = 0.072$, the $\varpi_2-\varpi_1$ resonance appears. 
    At $R = -0.0721$, the $\omega_2$ (inverse Kozai) resonance appears
    (dark blue and turquoise lobes centered on $\omega_2 = \pm 90^\circ$).
    At $R = 0$, $-0.0938$, and $-0.1373$, the $\varpi_2+\varpi_1-2\Omega_2$
    resonance manifests (this angle librates about $0^\circ$).
    These three resonances are accessed at inclinations $I \sim
    45^\circ$--75$^\circ$ (and at analogous retrograde inclinations
    that are not shown).
    The region at large $q_2$ for $R = -0.1373$ is empty because
    here the test particle locks into
    the $\Omega_2-\varpi_1$ resonance studied by \citet{naoz17},
    in which $\Omega_2-\varpi_1$ librates about $90^\circ$ and so does
    not trigger our sectioning criterion.}
    \label{fig:SOS_e1}
\end{figure*}

\begin{figure*}
	\includegraphics[width=\textwidth]{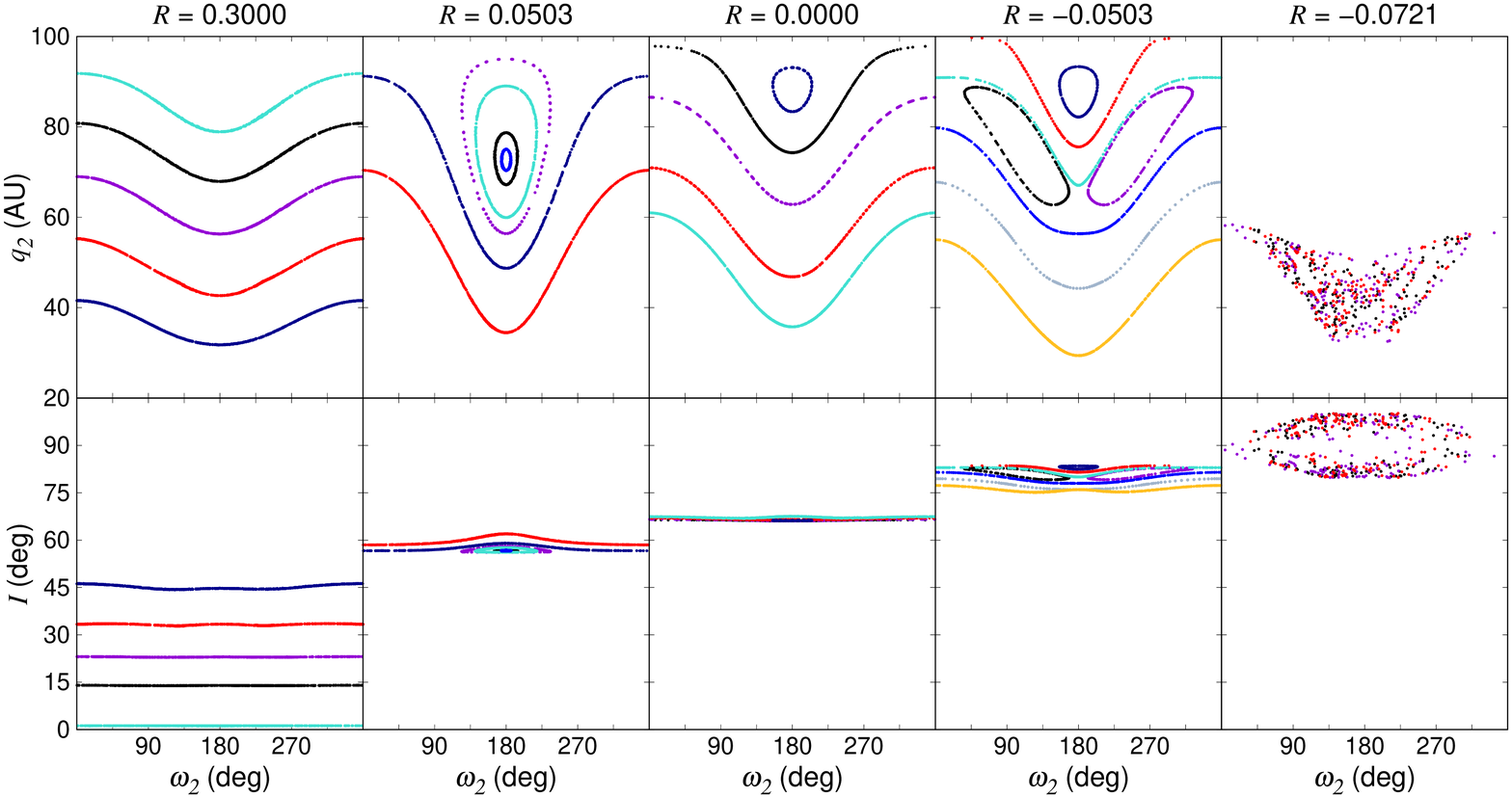}
    \caption{Same as Figure \ref{fig:SOS_e1} but for $e_1 = 0.3$.
The $\varpi_2-\varpi_1$ resonance appears at $R = 0.0503$;
the $\varpi_2 + \varpi_1 - 2\Omega_2$ resonance appears at
$R = 0$ and $-0.0503$; and the double-lobed 
inverse Kozai resonance appears at $R = -0.0503$.
Initial conditions used to make this figure are in Table
\ref{tab:SOS_e3}.}
    \label{fig:SOS_e3}
\end{figure*}

\begin{figure*}
	\includegraphics[width=\textwidth]{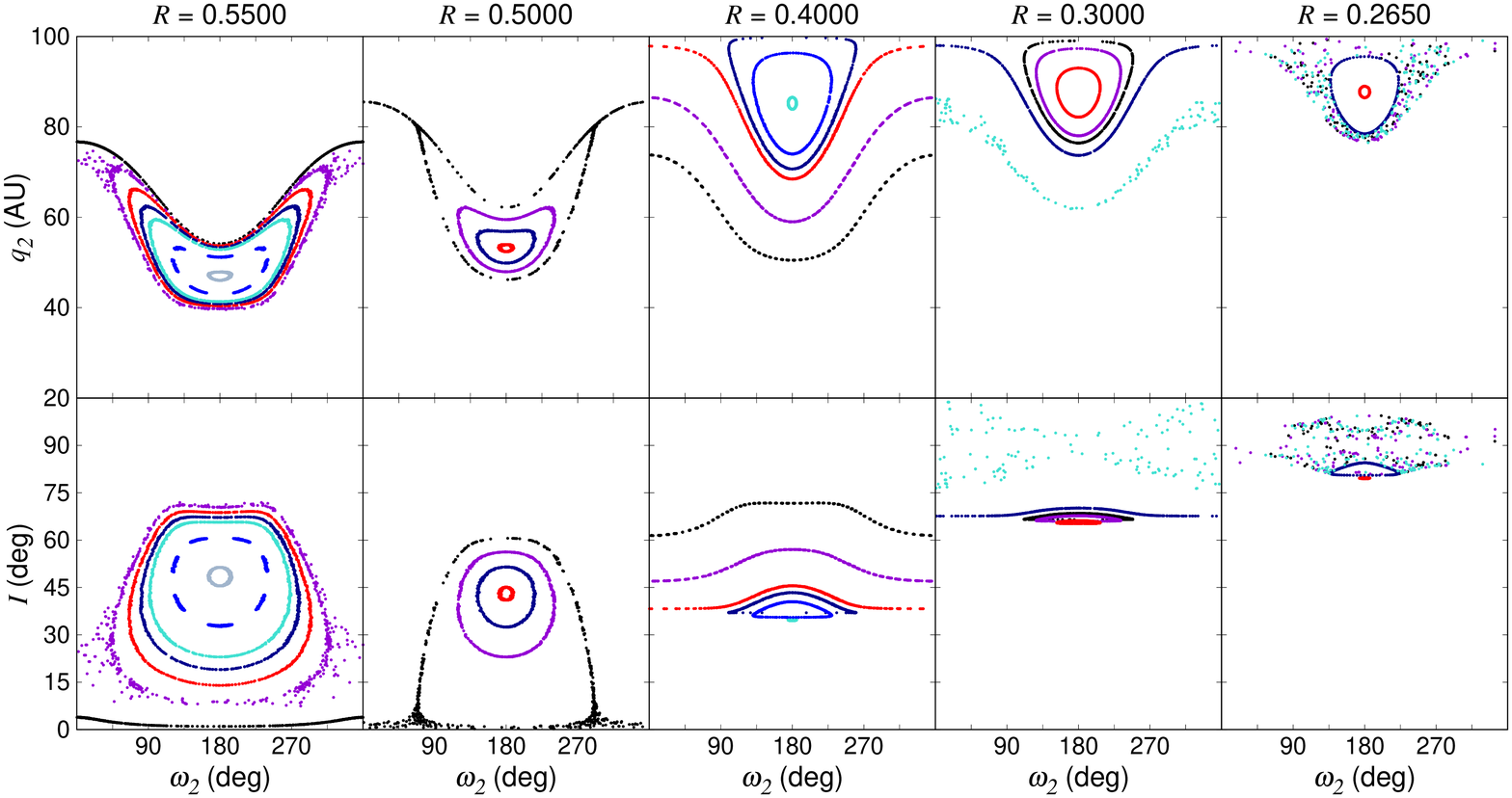}
    \caption{Same as Figure \ref{fig:SOS_e1} but for $e_1 = 0.7$.
In addition to the $\varpi_2-\varpi_1$ resonance at $R \leq 0.4$, a new
resonance appears at $R \geq 0.5$ for which
$\varpi_2-3\varpi_1+2\Omega_2$ $(= \omega_2 + 3 \Omega_2)$ librates
about $0^\circ$. 
This last resonance, however, 
is not found in full $N$-body integrations 
(by contrast to the other four resonances 
identified in this paper; see Figure \ref{fig:nbody}). 
Initial conditions used to make this figure
are in Table \ref{tab:SOS_e7}.}
    \label{fig:SOS_e7}
\end{figure*}

Since we are particularly interested in seeing how $\omega_2$ and
its quasi-conjugate $e_2$ behave,
we section using $\Omega_2$, plotting the particle's 
position in $q_2$-$\omega_2$ space and $I$-$\omega_2$ space
whenever $\Omega_2 = 180^\circ$ (with zero longitude defined
by $\varpi_1 = 0^\circ$), 
regardless of the sign of $\dot{\Omega}_2$. A conventional
SOS would select for $\dot{\Omega}_2$ of a single sign, but 
in practice there is no confusion; 
prograde orbits all have $\dot{\Omega}_2 < 0$ (see
equation \ref{eq:dOmegadt}) while retrograde orbits have
$\dot{\Omega}_2 > 0$; we focus for simplicity on prograde
orbits and capture a few retrograde branches
at the smallest values of $R$ (see the rightmost panels of
Figures \ref{fig:SOS_e1}--\ref{fig:SOS_e7}).
We have verified in a few cases
that the trajectories so plotted trace the maximum and minimum values
of $q_2$ and $I$; our SOS's contain the bounding
envelopes of the trajectories.

Figure \ref{fig:SOS_e1}
shows $\Omega_2$-SOS's for $e_1 = 0.1$ and a sequence of $R$'s 
(including those $R$ values used for Figures \ref{fig:inverse_Kozai_curves}
and \ref{fig:flip}).
At the most positive
$R$ (lowest $I$), the trajectories are regular, with small-amplitude
variations in $q_2$ and $I$. At more negative $R$ (larger $I$),
three strong resonances appear, each characterized by
substantial variations in $q_2$:
\begin{enumerate}

\item The first of these (appearing at $R = +0.0721$, second
panel from left)  is an ``apse-aligned'' resonance
for which $\varpi_2-\varpi_1$ librates about $0^\circ$ and
\begin{equation}
I (\varpi_2{\rm-}\varpi_1{\rm-res}) \simeq \arccos \left( \frac{+1 \pm \sqrt{6}}{5} \right) \simeq 46^\circ \, 
{\rm and} \, 107^\circ \,.
\end{equation}
At these inclinations, by equation (\ref{eq:dvarpidt}), 
$\left. d\varpi_2/dt\right|_{{\rm quad},e_1=0}=0$.\footnote{The
apse-aligned resonance identified here is at small $\alpha$ and large $I \simeq 46^\circ/107^\circ$,
but another apse-aligned resonance also exists
for orbits that are co-planar or nearly so (e.g., \citealt{wyatt99}). 
The latter can be found using
Laplace-Lagrange secular theory (e.g., \citealt{murray00}),
which does not expand in $\alpha$
but rather in eccentricity and inclination; it corresponds to a purely forced 
trajectory with no free oscillation. Laplace-Lagrange (read: low-inclination secular)
dynamics are well understood so we do not discuss them further.}

\item The second of the resonances ($R = -0.0721$, two lobes in the
middle panel)
is the inverse Kozai or $\omega_2$
resonance, appearing at $I (\omega_2{\rm-res}) \simeq 63^\circ$ and 
$117^\circ$, and for which $\omega_2$ librates about $\pm 90^\circ$
(Section \ref{sec:ikr}). 

\item The third resonance ($R = -0.0721$, $-0.0938$, and $-0.1373$;
middle, fourth, and fifth panels) appears at
\begin{equation}
I (\varpi_2{\rm+}\varpi_1{\rm-}2\Omega_2{\rm-res}) \simeq \arccos \left( \frac{-1 \pm \sqrt{6}}{5} \right) \simeq 73^\circ \, 
{\rm and} \, 134^\circ \,,
\end{equation}
inclinations for which $d(\varpi_2+\varpi_1-2\Omega_2)/dt = 0$ 
or equivalently $\dot{\omega}_2=\dot{\Omega}_2$ (equations
\ref{eq:domegadt} and \ref{eq:dOmegadt}). The resonant 
angle $\varpi_2 + \varpi_1 - 2\Omega_2$ $(= \omega_2-\Omega_2)$
librates about $0^\circ$.\footnote{The analogue of this
resonance for the interior test particle problem
has been invoked, together with other secular and mean-motion
resonant effects, in the context of Planet Nine and
Centaur evolution \citep{batygin17}.}

\end{enumerate}

For the above three resonances, we have verified that their
respective resonant arguments
($\varpi_2-\varpi_1$; $\omega_2$; $\varpi_2+\varpi_1-2\Omega_2$)
librate (see also Figure \ref{fig:nbody}), and have omitted their retrograde
branches from the SOS for simplicity.
The $\varpi_2+\varpi_1-2\Omega_2$ and $\varpi_2-\varpi_1$
resonances appear at octopole
order; they are associated with the first two terms
in the octopole disturbing function (\ref{eq:oct}), respectively.
The $\omega_2$ resonance is a hexadecapolar effect, as noted earlier.

The SOS for $e_1 = 0.3$ (Figure \ref{fig:SOS_e3}) reveals dynamics
qualitatively similar to $e_1 = 0.1$, but with larger amplitude variations
in $q_2$. We have verified in Figure \ref{fig:SOS_e3}
that the island of libration seen at $R = 0.0503$
is the $\varpi_2-\varpi_1$ resonance; that the islands near the top
of the panels for $R = 0$ and $-0.0503$ represent the
$\varpi_2+\varpi_1 - 2\Omega_2$ 
resonance; and that the two islands centered on $\omega_2 = \pm
90^\circ$ at $R = -0.0503$ represent the inverse Kozai resonance.

For both $e_1 = 0.3$ and $0.1$, chaos is more prevalent at
more negative $R$ / larger $I$. The chaotic
trajectories dip to periastron distances $q_2$ near $a_1 = 20$ AU, and
in Figure \ref{fig:SOS_e1} we show a few that actually cross orbits
with the perturber (the gray trajectory for $R = -0.0938$
is situated near the separatrix of the inverse Kozai resonance:
the two resonant lobes are seen in ghostly outline). 
The orbit-crossing behavior seen in Figure \ref{fig:SOS_e1}
occurs late in the test particle's evolution---in fact, at times longer
than the age of the universe for our parameter choices!
We nevertheless show these trajectories because the evolutionary 
timescales shorten and become realistic for smaller $a_1$ and $a_2$ 
(Section \ref{sec:prectime}). Unfortunately, no matter how we scale 
$a_1$ and $a_2$, the computational cost
of finding $N$-body counterparts to the orbit-crossing trajectories
of Figure \ref{fig:SOS_e1}
is necessarily expensive because the $N$-body timestep scales with the
orbital period of the interior body; $N$-body tests of these
particular trajectories are deferred to future work.

At $e_1 = 0.7$ (Figure \ref{fig:SOS_e7}) we find, in addition to the
$\varpi_2-\varpi_1$ resonance at $R \leq 0.4$, a new resonance at $R \geq 0.5$.
For this latter resonance,
$\varpi_2 - 3 \varpi_1 + 2 \Omega_2 = \omega_2 + 3 \Omega_2$
librates about $0^\circ$.
Although this resonance is found at octopole order---it is embodied in
the fourth term in equation (\ref{eq:oct})---we found by
experimentation that eliminating the hexadecapole contribution to the
disturbing function removes the test particle from this resonance (for
the same initial conditions as shown in Figure \ref{fig:SOS_e7}).
Evidently the hexadecapole potential helps to enforce
$\dot{\omega}_2 = - 3 \dot{\Omega}_2$ so that this octopole resonance
can be activated. Remarkably, this $\varpi_2-3\varpi_1+2\Omega_2$
resonance enables the test particle to cycle between a nearly (but not
exactly) co-planar orbit to one inclined by
$\sim$60--70$^\circ$, while having its eccentricity $e_2$
vary between $\sim$0.2--0.6. 
We will see in Section \ref{sec:nbody}, however,
that a full $N$-body treatment mutes the effects of this resonance.



\subsection{$N$-Body Tests} \label{sec:nbody}
\begin{figure*}
	\includegraphics[width=\textwidth]{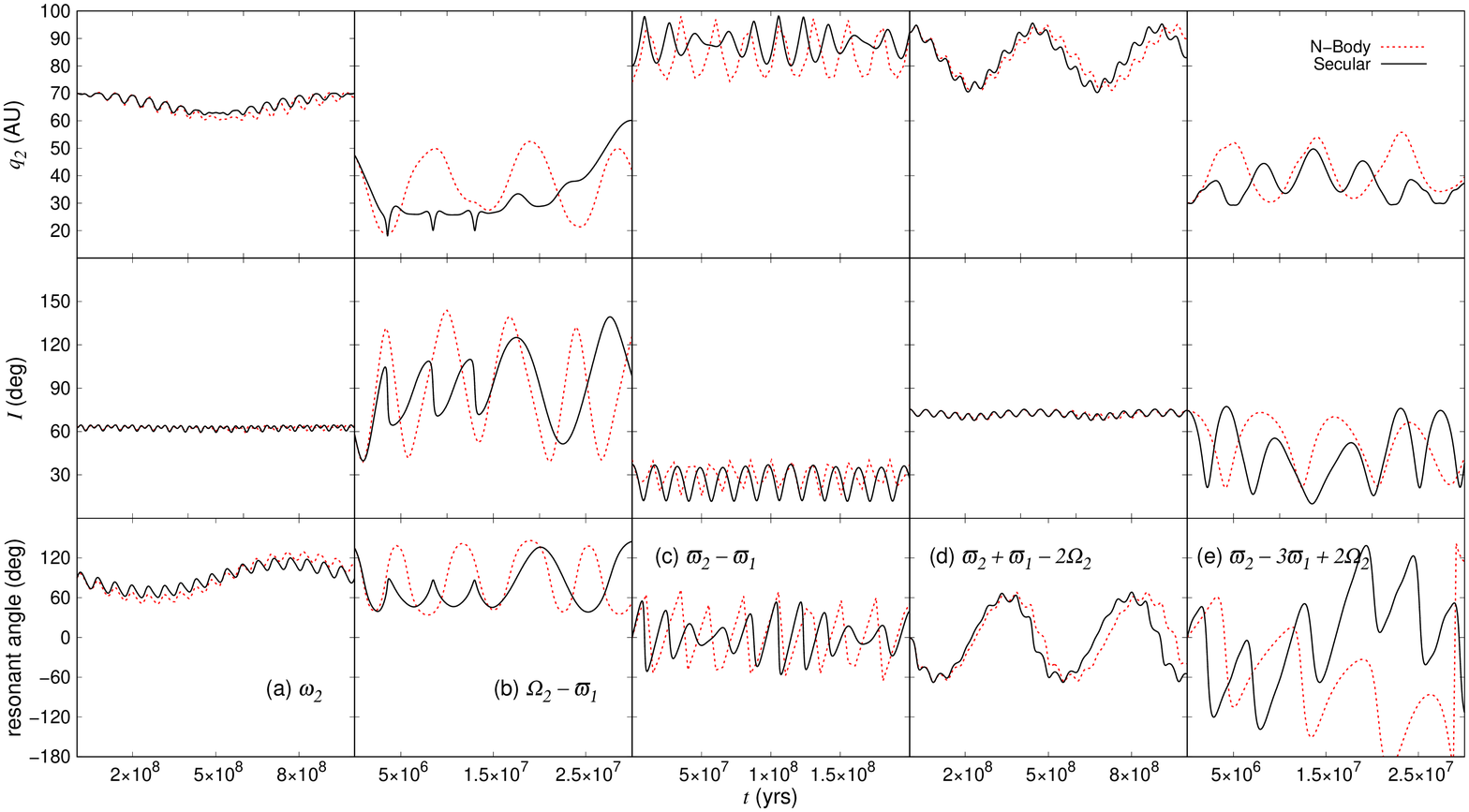}
\caption{Comparison of $N$-body (dashed red)
vs. secular (solid black) integrations.
Initial conditions, summarized in Table \ref{tab:nbody},
are chosen to lock the test particle into the $\omega_2$,
$\Omega_2-\varpi_1$, $\varpi_2-\varpi_1$, and $\varpi_2+\varpi_1-2\Omega_2$ resonances
(panels a through d). We failed to obtain a lock for the
$\varpi_2 - 3 \varpi_1 +2\Omega_2 = \omega_2+3\Omega_2$ resonance in our $N$-body calculations
and offer an instead an $N$-body trajectory that
comes close to librating
($\dot{\omega}_2 \approx -2.7 \dot{\Omega}_2$; panel e),
together with its secular counterpart which does
appear to librate.
}
    \label{fig:nbody}
\end{figure*}

Having identified five resonances in the above surveys,
we test how robust they are using $N$-body integrations.
We employ the \texttt{WHFast} symplectic
integrator (\citealt{rein15}; \citealt{wisdom91}), part of the
\texttt{REBOUND} package (\citealt{rein12}), adopting
timesteps between 0.1--0.25 yr. 
Initial conditions (inputted for the $N$-body experiments
as Jacobi elements, together with initial true anomalies
$f_1 = 0^\circ$ and $f_2 = 180^\circ$)
were drawn from the above surveys
with the goal of finding resonant
libration at as large a perturber eccentricity $e_1$ as possible.
In Figure \ref{fig:nbody} we verify that the
$\omega_2$, $\Omega_2-\varpi_1$, $\varpi_2-\varpi_1$, and
$\varpi_2 +\varpi_2-2\Omega_2$
resonances survive a full $N$-body treatment when $e_1$
is as high as $0.1$, $0.7$, $0.7$, and $0.1$, respectively
(see also Figure \ref{fig:gallardo_time}).
Table \ref{tab:nbody} records the initial conditions.

We were unable in $N$-body calculations
to lock the test particle
into the $\varpi_2-3\varpi_1+2\Omega_2$ $(= \omega_2 + 3\Omega_2)$ resonance,
despite exploring the parameter
space in the vicinity where we found it in
the secular surfaces of section.
This is unsurprising insofar as we had found
this particular
resonance to depend on both octopole and hexadecapolar effects
at the largest perturber eccentricity
tested, $e_1 = 0.7$; at such a high eccentricity, effects even higher
order than hexadecapole are likely to be significant, and it appears
from our $N$-body calculations that they are, preventing a resonant lock.
We show in Figure \ref{fig:nbody} an $N$-body 
trajectory that comes close to
being in
this
resonance (on average,
$\dot{\omega}_2 \approx -2.7 \dot{\Omega}_2$). Although the
inclination does not vary as dramatically
as in the truncated secular evolution, it can still cycle
between $\sim$$20^\circ$ and $70^\circ$.

The agreement between the $N$-body and secular integrations shown
in Figure \ref{fig:nbody} is good, qualitatively and even quantitatively
in some cases. We emphasize that these trajectories have not been cherry-picked
to display such agreement; the initial conditions were drawn from the
preceding surveys for the purpose of testing which resonances
survive an $N$-body treatment. In the cases of the $\Omega_2-\varpi_1$
and $\varpi_2-\varpi_1$ resonances, the secular trajectories
show amplitude modulation/beating not seen in their
$N$-body counterparts. Similar behavior was reported
by \citet[][see their Figure 12]{naoz17}. A broader continuum
of forcing frequencies must be present at our standard value of
$\alpha = 0.2$ than is captured by our hex-limited treatment;
certainly we obtain better agreement with $N$-body calculations
at lower values of $\alpha$ (as we have explicitly verified by testing, e.g., 
$\alpha = 0.05$ for the parameters of Figure \ref{fig:nbody}c).

\section{Summary}\label{sec:conclude}

We have surveyed numerically the dynamics
of an external test particle in the restricted,
secular, three-body problem. We wrote down the secular potential
of an internal perturber to hexadecapolar order (where
the expansion parameter is the ratio of
semimajor axes of the internal and external bodies,
$\alpha = a_1/a_2 < 1$) 
by adapting the disturbing function for an external perturber as
derived by 
\citet[][Y03]{yokoyama03}. In making this adaptation,
we corrected a misprint in the hexadecapolar potential 
of Y03 (M.~\'Cuk 2017, personal communication).
Our numerical survey was conducted at fixed $\alpha = 0.2$,
the largest value we thought might still be captured
by a truncated secular expansion (lower values generally do better).

Inclination variations for an external test particle
can be dramatic when the eccentricity of the internal
perturber $e_1$ is non-zero. The variations in mutual
inclination $I$ are effected by a
quadrupole resonance for which $\Omega_2$, the test
particle's longitude of ascending node (referenced to the orbit plane
of the perturber, whose periapse is at longitude $\varpi_1$),
librates about $\varpi_1 \pm 90^\circ$. Within this $\Omega_2-\varpi_1$
resonance, the test particle's
orbit flips (switches from prograde to retrograde). Flipping
is easier---i.e., the minimum
$I$ for which flipping is possible decreases---with increasing
$e_1$. All of this inclination behavior was described by 
Naoz et al.~(\citeyear{naoz17}; see also \citealt{verrier09}
and \citealt{farago10})
and we have confirmed these essentially quadrupolar
results here. 

Eccentricity variations for an external test particle
rely on octopole or higher-level effects (at the quadrupole
level of approximation, the test particle
eccentricity $e_2$ is strictly constant).
When $e_1 = 0$, octopole effects vanish, and
the leading-order resonance able to produce
eccentricity variations is the hexadecapolar
``inverse Kozai'' resonance
in which the test particle's argument of periastron
$\omega_2$ librates about $\pm 90^\circ$ \citep{gallardo12}.
The resonance demands rather high inclinations,
$I \simeq 63^\circ$ or $117^\circ$.
By comparison to its conventional Kozai counterpart which exists
at quadrupole order, the hexadecapolar
inverse Kozai resonance is more restricted
in scope: it exists only over a narrow range of
$J_z = \sqrt{1-e_2^2}\cos I$ for a given $\alpha$,
and produces eccentricity variations on the order of
$\Delta e_2 \simeq 0.2$. For suitable $J_z$
it can, however, lead to orbit-crossing with the perturber.
In our truncated secular treatment,
we found the inverse Kozai resonance
to persist up to perturber eccentricities of $e_1 = 0.3$;
in $N$-body experiments, we found the resonance only
up to $e_1 = 0.1$.
At higher $e_1$, the hexadecapolar
resonance seems to disappear,
overwhelmed by octopole effects.

Surfaces of section made for $e_1 \neq 0$
and $\Omega_2 = 180^\circ$
revealed two
octopole resonances
characterized by stronger eccentricity variations of
$\Delta e_2$ up to 0.5. The resonant angles are the apsidal difference 
$\varpi_2-\varpi_1$, which librates
about $0^\circ$, and 
$\varpi_2+\varpi_1-2\Omega_2$, which also librates about $0^\circ$.
The $\varpi_2-\varpi_1$ and $\varpi_2+\varpi_1-2\Omega_2$ resonances
are like the inverse Kozai resonance in that they
also require large inclinations,
$I \simeq 46^\circ$/$107^\circ$ and $73^\circ/134^\circ$,
respectively. 
The apse-aligned $\varpi_2-\varpi_1$ resonance survives
full $N$-body integrations up to $e_1 = 0.7$;
the $\varpi_2+\varpi_1-2\Omega_2$ resonance survives up to $e_1 = 0.1$.
At large $e_1$, the requirement on $I$ for the $\varpi_2-\varpi_1$ resonance
lessens to about $\sim$20$^\circ$.

We outlined two rough, qualitative trends: (1)
the larger $e_1$ is, the more
the eccentricity and inclination of the test particle can vary;
and (2) the more polar the test particle orbit (i.e.,
the closer $I$ is to $90^\circ$), the more chaotic its evolution.

In some high-inclination trajectories---near the separatrix
of the inverse Kozai resonance, for example---test particle
periastra could be lowered from large distances to near the perturber.
These secular channels of transport
need to be confirmed with $N$-body tests.

This paper is but an initial reconnaissance of the external
test particle problem. How the various resonances we have identified
may have operated in practice to shape actual planetary/star
systems is left for future study. In addition to more
$N$-body tests, we also need to explore the effects
of general relativity (GR). For our chosen parameters,
GR causes the periapse of the perturber to precess
at a rate that is typically several hundreds of times slower
than the rate at which the test particle's node precesses.
Such an additional apsidal precession is not expected to affect
our results materially; still, a check should be made.
A way to do that comprehensively is
to re-compute our surfaces of section with GR.

\section*{Acknowledgements}

We are grateful to Edgar Knobloch and Matthias Reinsch for teaching
Berkeley's upper-division mechanics course Physics 105,
and for connecting BV with EC.
This work was supported by a Berkeley Excellence Account for Research
and the NSF.
Matija \'Cuk alerted us to the misprint in Y03
and provided insights that were most helpful.
We thank Smadar Naoz for an encouraging and constructive
referee's report; Konstantin Batygin, Alexandre Correira, Bekki
Dawson, 
Eve Lee, Yoram Lithwick,
and Renu Malhotra for useful discussions and feedback;
and Daniel Tamayo for teaching us how to use \texttt{REBOUND}.




\bibliographystyle{mnras}
\bibliography{inversekozai} 

\begin{thebibliography}{}
\makeatletter
\relax
\def\mn@urlcharsother{\let\do\@makeother \do\$\do\&\do\#\do\^\do\_\do\%\do\~}
\def\mn@doi{\begingroup\mn@urlcharsother \@ifnextchar [ {\mn@doi@}
  {\mn@doi@[]}}
\def\mn@doi@[#1]#2{\def\@tempa{#1}\ifx\@tempa\@empty \href
  {http://dx.doi.org/#2} {doi:#2}\else \href {http://dx.doi.org/#2} {#1}\fi
  \endgroup}
\def\mn@eprint#1#2{\mn@eprint@#1:#2::\@nil}
\def\mn@eprint@arXiv#1{\href {http://arxiv.org/abs/#1} {{\tt arXiv:#1}}}
\def\mn@eprint@dblp#1{\href {http://dblp.uni-trier.de/rec/bibtex/#1.xml}
  {dblp:#1}}
\def\mn@eprint@#1:#2:#3:#4\@nil{\def\@tempa {#1}\def\@tempb {#2}\def\@tempc
  {#3}\ifx \@tempc \@empty \let \@tempc \@tempb \let \@tempb \@tempa \fi \ifx
  \@tempb \@empty \def\@tempb {arXiv}\fi \@ifundefined
  {mn@eprint@\@tempb}{\@tempb:\@tempc}{\expandafter \expandafter \csname
  mn@eprint@\@tempb\endcsname \expandafter{\@tempc}}}

\bibitem[\protect\citeauthoryear{Bailey, Chambers  \& Hahn}{Bailey
  et~al.}{1992}]{bailey92}
Bailey M.~E.,  Chambers J.~E.,   Hahn G.,  1992, Astronomy and Astrophysics,
  257, 315

\bibitem[\protect\citeauthoryear{{Batygin} \& {Morbidelli}}{{Batygin} \&
  {Morbidelli}}{2017}]{batygin17}
{Batygin} K.,  {Morbidelli} A.,  2017, preprint, \href
  {http://adsabs.harvard.edu/abs/2017arXiv171001804B} {} (\mn@eprint {arXiv}
  {1710.01804})

\bibitem[\protect\citeauthoryear{Carruba}{Carruba}{2002}]{carruba02}
Carruba V.,  2002, Icarus, 158, 434

\bibitem[\protect\citeauthoryear{{Dawson} \& {Chiang}}{{Dawson} \&
  {Chiang}}{2014}]{dawson14}
{Dawson} R.~I.,  {Chiang} E.,  2014, \mn@doi [Science]
  {10.1126/science.1256943}, \href
  {http://adsabs.harvard.edu/abs/2014Sci...346..212D} {346, 212}

\bibitem[\protect\citeauthoryear{{Farago} \& {Laskar}}{{Farago} \&
  {Laskar}}{2010}]{farago10}
{Farago} F.,  {Laskar} J.,  2010, \mn@doi [\mnras]
  {10.1111/j.1365-2966.2009.15711.x}, \href
  {http://adsabs.harvard.edu/abs/2010MNRAS.401.1189F} {401, 1189}

\bibitem[\protect\citeauthoryear{Gallardo, Hugo  \& Pais}{Gallardo
  et~al.}{2012}]{gallardo12}
Gallardo T.,  Hugo G.,   Pais P.,  2012, Icarus, 220, 392

\bibitem[\protect\citeauthoryear{{Katz}, {Dong}  \& {Malhotra}}{{Katz}
  et~al.}{2011}]{katz11}
{Katz} B.,  {Dong} S.,   {Malhotra} R.,  2011, \mn@doi [Physical Review
  Letters] {10.1103/PhysRevLett.107.181101}, \href
  {http://adsabs.harvard.edu/abs/2011PhRvL.107r1101K} {107, 181101}

\bibitem[\protect\citeauthoryear{Kozai}{Kozai}{1962}]{kozai62}
Kozai Y.,  1962, The Astronomical Journal, 67, 591

\bibitem[\protect\citeauthoryear{Kushnir, Katz, Dong, Livne  \&
  Fern{\'a}ndez}{Kushnir et~al.}{2013}]{kushnir13}
Kushnir D.,  Katz B.,  Dong S.,  Livne E.,   Fern{\'a}ndez R.,  2013, The
  Astrophysical Journal, 778, L37

\bibitem[\protect\citeauthoryear{Lee \& Chiang}{Lee \& Chiang}{2016}]{lee16}
Lee E.~J.,  Chiang E.,  2016, The Astrophysical Journal, 827, 125

\bibitem[\protect\citeauthoryear{Lidov}{Lidov}{1962}]{lidov62}
Lidov M.~L.,  1962, Planetary and Space Science, 9, 719

\bibitem[\protect\citeauthoryear{Lithwick \& Naoz}{Lithwick \&
  Naoz}{2011}]{lithwick11}
Lithwick Y.,  Naoz S.,  2011, The Astrophysical Journal, 742, 94

\bibitem[\protect\citeauthoryear{{Macintosh} et~al.,}{{Macintosh}
  et~al.}{2015}]{macintosh15}
{Macintosh} B.,  et~al., 2015, \mn@doi [Science] {10.1126/science.aac5891},
  \href {http://adsabs.harvard.edu/abs/2015Sci...350...64M} {350, 64}

\bibitem[\protect\citeauthoryear{{Murray} \& {Dermott}}{{Murray} \&
  {Dermott}}{2000}]{murray00}
{Murray} C.~D.,  {Dermott} S.~F.,  2000, {Solar System Dynamics}.
''Cambridge University Press''

\bibitem[\protect\citeauthoryear{{Naoz}}{{Naoz}}{2016}]{naoz16}
{Naoz} S.,  2016, \mn@doi [\araa] {10.1146/annurev-astro-081915-023315}, \href
  {http://adsabs.harvard.edu/abs/2016ARA%26A..54..441N} {54, 441}

\bibitem[\protect\citeauthoryear{Naoz, Farr, Lithwick, Rasio  \&
  Teyssandier}{Naoz et~al.}{2011}]{naoz11}
Naoz S.,  Farr W.~M.,  Lithwick Y.,  Rasio F.~A.,   Teyssandier J.,  2011,
  Nature, 473, 187

\bibitem[\protect\citeauthoryear{Naoz, Li, Zanardi, de El{\'\i}a  \&
  Di~Sisto}{Naoz et~al.}{2017}]{naoz17}
Naoz S.,  Li G.,  Zanardi M.,  de El{\'\i}a G.~C.,   Di~Sisto R.~P.,  2017, The
  Astronomical Journal, 154, 18

\bibitem[\protect\citeauthoryear{Nesvold, Naoz, Vican  \& Farr}{Nesvold
  et~al.}{2016}]{nesvold16}
Nesvold E.~R.,  Naoz S.,  Vican L.,   Farr W.~M.,  2016, The Astrophysical
  Journal, 826, 19

\bibitem[\protect\citeauthoryear{Nesvorny, Alvarellos, Dones  \&
  Levison}{Nesvorny et~al.}{2003}]{nesvorny03}
Nesvorny D.,  Alvarellos J. L.~A.,  Dones L.,   Levison H.~F.,  2003, The
  Astronomical Journal, 126, 398

\bibitem[\protect\citeauthoryear{Pearce \& Wyatt}{Pearce \&
  Wyatt}{2014}]{pearce14}
Pearce T.~D.,  Wyatt M.~C.,  2014, Monthly Notices of the Royal Astronomical
  Society, 443, 2541

\bibitem[\protect\citeauthoryear{{Rein} \& {Liu}}{{Rein} \&
  {Liu}}{2012}]{rein12}
{Rein} H.,  {Liu} S.-F.,  2012, \mn@doi [\aap] {10.1051/0004-6361/201118085},
  \href {http://adsabs.harvard.edu/abs/2012A%26A...537A.128R} {537, A128}

\bibitem[\protect\citeauthoryear{{Rein} \& {Tamayo}}{{Rein} \&
  {Tamayo}}{2015}]{rein15}
{Rein} H.,  {Tamayo} D.,  2015, \mn@doi [\mnras] {10.1093/mnras/stv1257}, \href
  {http://adsabs.harvard.edu/abs/2015MNRAS.452..376R} {452, 376}

\bibitem[\protect\citeauthoryear{Sheppard \& Trujillo}{Sheppard \&
  Trujillo}{2016}]{sheppard16}
Sheppard S.~S.,  Trujillo C.,  2016, The Astronomical Journal, 152, 221

\bibitem[\protect\citeauthoryear{Silsbee \& Tremaine}{Silsbee \&
  Tremaine}{2016}]{silsbee16}
Silsbee K.,  Tremaine S.,  2016, The Astronomical Journal, 152, 103

\bibitem[\protect\citeauthoryear{Silsbee \& Tremaine}{Silsbee \&
  Tremaine}{2017}]{silsbee17}
Silsbee K.,  Tremaine S.,  2017, The Astrophysical Journal, 836, 39

\bibitem[\protect\citeauthoryear{Thomas \& Morbidelli}{Thomas \&
  Morbidelli}{1996}]{thomas96}
Thomas F.,  Morbidelli A.,  1996, Celestial Mechanics, 64, 209

\bibitem[\protect\citeauthoryear{Tremaine \& Yavetz}{Tremaine \&
  Yavetz}{2014}]{tremaine14}
Tremaine S.,  Yavetz T.~D.,  2014, American Journal of Physics, 82, 769

\bibitem[\protect\citeauthoryear{{Verrier} \& {Evans}}{{Verrier} \&
  {Evans}}{2008}]{verrier08}
{Verrier} P.~E.,  {Evans} N.~W.,  2008, \mn@doi [\mnras]
  {10.1111/j.1365-2966.2008.13854.x}, \href
  {http://adsabs.harvard.edu/abs/2008MNRAS.390.1377V} {390, 1377}

\bibitem[\protect\citeauthoryear{{Verrier} \& {Evans}}{{Verrier} \&
  {Evans}}{2009}]{verrier09}
{Verrier} P.~E.,  {Evans} N.~W.,  2009, \mn@doi [\mnras]
  {10.1111/j.1365-2966.2009.14446.x}, \href
  {http://adsabs.harvard.edu/abs/2009MNRAS.394.1721V} {394, 1721}

\bibitem[\protect\citeauthoryear{{Wisdom} \& {Holman}}{{Wisdom} \&
  {Holman}}{1991}]{wisdom91}
{Wisdom} J.,  {Holman} M.,  1991, \mn@doi [\aj] {10.1086/115978}, \href
  {http://adsabs.harvard.edu/abs/1991AJ....102.1528W} {102, 1528}

\bibitem[\protect\citeauthoryear{Wu \& Murray}{Wu \& Murray}{2003}]{wu03}
Wu Y.,  Murray N.,  2003, The Astrophysical Journal, 589, 605

\bibitem[\protect\citeauthoryear{Wyatt, Dermott, Telesco, Fisher, Grogan,
  Holmes  \& Pi{\~n}a}{Wyatt et~al.}{1999}]{wyatt99}
Wyatt M.~C.,  Dermott S.~F.,  Telesco C.~M.,  Fisher R.~S.,  Grogan K.,  Holmes
  E.~K.,   Pi{\~n}a R.~K.,  1999, The Astrophysical Journal, 527, 918

\bibitem[\protect\citeauthoryear{Yokoyama, Santos, Cardin  \& Winter}{Yokoyama
  et~al.}{2003}]{yokoyama03}
Yokoyama T.,  Santos M.~T.,  Cardin G.,   Winter O.~C.,  2003, Astronomy and
  Astrophysics, 401, 763

\bibitem[\protect\citeauthoryear{Zanardi, de El{\'\i}a, Di~Sisto, Naoz, Li,
  Guilera  \& Brunini}{Zanardi et~al.}{2017}]{zanardi17}
Zanardi M.,  de El{\'\i}a G.~C.,  Di~Sisto R.~P.,  Naoz S.,  Li G.,  Guilera
  O.~M.,   Brunini A.,  2017, Astronomy and Astrophysics, 605, A64

\makeatother
\end{thebibliography}




\appendix

\section{Initial Conditions}
\begin{table}
    \caption{Initial conditions for Figure \ref{fig:inverse_Kozai_curves}
    ($R = -0.05032$).}\label{tab:inverse_Kozai_curves}
    \begin{tabular}{r|r|r|r|r}
\hline\hline 
\multicolumn{1}{c}{$e_2$} & \multicolumn{1}{c}{$\Omega_2$} & \multicolumn{1}{c}{$\varpi_2$} & \multicolumn{1}{c}{$I$} & \multicolumn{1}{c}{$J_{z,\rm init}$}\\
    & \multicolumn{1}{c}{(rad)} & \multicolumn{1}{c}{(rad)} & \multicolumn{1}{c}{(rad)} \\
\hline\hline
\multicolumn{5}{c}{Center Panel ($e_1=0.1$, $\varpi_1=0$)}\\
\hline
0.2110 & 0 & 1.5708 & 1.1170 & 0.4285\\
0.2110 & 0 & 1.5708 & 2.0246 & -0.4285\\
0.4165 &  0 & 1.5708 & 1.0821 & 0.4268\\
0.4165 &  0 & 1.5708 & 2.0595 & -0.4268\\
0.6738 & 0 & 0 & 1.0123 & 0.3916\\ 
0.6738 & 0 & 0 & 2.1293 & -0.3916\\
\hline\hline 
\multicolumn{5}{c}{Bottom Panel ($e_1=0.3$, $\varpi_1=0$)}\\
\hline
0.1094 & 0 & 1.5708 & 1.4486 & 0.1212\\
0.1094 & 0 & 1.5708 & 1.6930 & -0.1212\\
0.5575 & 0 & 0 & 1.3264 & 0.2009\\
0.5575 & 0 & 0 & 1.8151 & -0.2009\\
0.7061 & 0 & 0 & 1.3264 & 0.1713\\
0.7061 & 0 & 0 & 1.8151 & -0.1713\\
\hline \hline
    \end{tabular}
\end{table}

\begin{table}
\caption{Initial conditions for Figure \ref{fig:SOS_e1} ($e_1=0.1$, $\varpi_1 = 0$).}\label{tab:SOS_e1}
\begin{tabular}{|r|r|r|r|r|} 
\hline\hline 
\multicolumn{1}{c}{color} & \multicolumn{1}{c}{$e_2$} & \multicolumn{1}{c}{$\Omega_2$} & \multicolumn{1}{c}{$\varpi_2$} & \multicolumn{1}{c}{$I$} \\
 & & \multicolumn{1}{c}{(rad)} & \multicolumn{1}{c}{(rad)} & \multicolumn{1}{c}{(rad)} \\
\hline\hline 
\multicolumn{5}{c}{$R = 0.3000$}\\
\hline 
 black & 0.3189 & 0.0000 & 0.0000 & 0.0175\\
 violet & 0.3974 & 0.0000 & 0.0000 & 0.2443\\
 red & 0.5421 & 0.0000 & 0.0000 & 0.0454 \\
 dark blue & 0.4656 & -0.7854 & 0.0000 & 0.3491\\
 turquoise & 0.6604 & -0.7854 & 0.0000 & 0.5934\\
\hline\hline 
\multicolumn{5}{c}{$R = 0.0721$}\\
\hline 
 black & 0.0678 & -0.7854 & 0.0000 & 0.7592\\ 
 violet & 0.1894 & -0.7854 & 0.0000 & 0.7679\\ 
 red & 0.3685 & -0.7854 & 0.0000 & 0.7941 \\
 dark blue & 0.4556 & -0.7854 & 0.0000 & 0.8116 \\
 turquoise & 0.6036 & 0.0000 & 0.0000 & 0.8465 \\
 blue & 0.0957 & -1.5708 & 0.0000 & 0.7505 \\
 gray & 0.6683 & -1.5708 & 0.0000 & 0.8552\\
\hline\hline 
\multicolumn{5}{c}{$R = -0.0721$}\\
\hline 
 black & 0.0072 & -1.5708 & 0.0000 & 1.1519\\ 
 violet & 0.5842 & 0.0000 & 0.0000 & 1.0821\\ 
 red & 0.2535 & 0.0000 & 0.3491 & 1.1868 \\
 dark blue & 0.4882 & 0.0000 & 0.9076 & 1.1170 \\
 turquoise & 0.4882 & -3.1416 & 0.9076 & 1.1170 \\
 blue & 0.3832 & 0.0000 & 1.3963 & 1.1519 \\
 gray & 0.1479 & 0.0000 & 1.4661 & 1.2042\\
 yellow & 0.2485 & 0.0000 & 1.5359 & 1.1868 \\
\hline \hline 
\multicolumn{5}{c}{$R = -0.0938$}\\
\hline 
 black & 0.0787 & 0.0000 & 0.0000 & 1.3177\\ 
 violet & 0.1563 & -0.7854 & 0.0000 & 1.2654\\ 
 red & 0.2028 & -0.7854 & 0.0000 & 1.2566 \\
 dark blue & 0.2733 & -0.7854 & 0.0000 & 1.2392 \\
 turquoise & 0.5353 & 0.0000 & 0.1396 & 1.1519 \\
 blue & 0.3088 & 0.0000 & 0.4189 & 1.2566 \\
 gray & 0.5957 & 0.0000 & 0.6981 & 1.1170\\
 yellow & 0.4617 & 0.0000 & 1.3963 & 1.1868 \\
\hline \hline 
\multicolumn{5}{c}{$R = -0.1373$}\\
\hline 
 black & 0.3691 & 0.0000 & 0.0000 & 1.4835\\ 
 violet & 0.3500 & 0.0000 & 0.3491 & 1.5533\\ 
 red & 0.4684 & 0.0000 & 0.6283 & 1.3439 \\
 dark blue & 0.3565 & 0.0000 & 0.8378 & 1.5185 \\
 turquoise & 0.3326 & 0.0000 & 0.2793 & 1.0996 \\
 blue & 0.5176 & 0.0000 & 1.2566 & 1.2915 \\
\hline \hline 
\end{tabular}
\end{table}

\begin{table}
\caption{Initial conditions for Figure \ref{fig:SOS_e3}
($e_1=0.3$, $\varpi_1 = 0$).}\label{tab:SOS_e3}
\begin{tabular}{|r|r|r|r|r|} 
\hline\hline
\multicolumn{1}{c}{color} & \multicolumn{1}{c}{$e_2$} & \multicolumn{1}{c}{$\Omega_2$} & \multicolumn{1}{c}{$\varpi_2$} & \multicolumn{1}{c}{$I$} \\
 & & \multicolumn{1}{c}{(rad)} & \multicolumn{1}{c}{(rad)} & \multicolumn{1}{c}{(rad)} \\
\hline\hline
\multicolumn{5}{c}{$R = 0.3000$}\\
\hline
 black & 0.3207 & 0.0000 & 0.0000 & 0.2443\\
 violet & 0.4371 & 0.0000 & 0.0000 & 0.4014\\
 red & 0.5534 & -1.5708 & 0.0000 & 0.4712 \\
 dark blue & 0.6588 & -1.5708 & 0.0000 & 0.6283\\
 turquoise & 0.2110 & -1.5708 & 0.0000 & 0.0175\\
\hline\hline
\multicolumn{5}{c}{$R = 0.0503$}\\
\hline
 black & 0.3279 & 0.0000 & 0.0000 & 0.9948\\ 
 violet & 0.4362 & 0.0000 & 0.0000 & 1.0123\\ 
 red & 0.6553 & 0.0000 & 0.0000 & 1.0821 \\
 dark blue & 0.5130 & 0.0000 & 0.0000 & 1.0297 \\
 turquoise & 0.0817 & -0.7854 & 0.0000 & 0.8378 \\
 blue & 0.2608 & -0.7854 & 0.0000 & 0.8552 \\
\hline\hline
\multicolumn{5}{c}{$R = 0.0000$}\\
\hline
 black & 0.2127 & -0.7854 & 0.0000 & 0.9687\\ 
 violet & 0.3300 & -0.7854 & 0.0000 & 0.9774\\ 
 red & 0.4942 & -0.7854 & 0.0000 & 0.9948 \\
 dark blue & 0.0985 & -1.5708 & 0.0000 & 0.8465 \\
 turquoise & 0.5784 & -1.5708 & 0.0000 & 0.9163 \\
\hline \hline
\multicolumn{5}{c}{$R = -0.0503$}\\
\hline
 black & 0.3380 & -0.7854 & 0.0000 & 1.0996\\ 
 violet & 0.3380 & -3.9270 & 0.0000 & 1.0996\\ 
 red & 0.2444 & 0.0000 & 0.0000 & 1.4224 \\
 dark blue & 0.0667 & 0.0000 & 0.0000 & 1.4573 \\
 turquoise & 0.3295 & 0.0000 & 0.0000 & 1.3963 \\
 blue & 0.4364 & 0.0000 & 0.0000 & 1.3614 \\
 gray & 0.5576 & 0.0000 & 0.0000 & 1.3265\\
 yellow & 0.7061 & 0.0000 & 0.0000 & 1.3265 \\
\hline \hline
\multicolumn{5}{c}{$R = -0.0721$}\\
\hline
 black & 0.4081 & -0.7854 & 0.0000 & 1.1519\\ 
 violet & 0.4937 & 0.0000 & 0.0000 & 1.5184\\ 
 red & 0.5541 & 0.0000 & 0.0000 & 1.4486 \\
\hline \hline
\end{tabular}
\end{table}

\begin{table}
\caption{Initial conditions for Figure \ref{fig:SOS_e7}
($e_1=0.7$, $\varpi_1 = 0$).}\label{tab:SOS_e7}
\begin{tabular}{|r|r|r|r|r|} 
\hline\hline
\multicolumn{1}{c}{color} & \multicolumn{1}{c}{$e_2$} & \multicolumn{1}{c}{$\Omega_2$} & \multicolumn{1}{c}{$\varpi_2$} & \multicolumn{1}{c}{$I$} \\
 & & \multicolumn{1}{c}{(rad)} & \multicolumn{1}{c}{(rad)} & \multicolumn{1}{c}{(rad)} \\
\hline\hline
\multicolumn{5}{c}{$R = 0.5500$}\\
\hline
 black & 0.4587 & 0.0000 & 0.0000 & 0.0175\\
 violet & 0.4629 & 0.0000 & 0.0000 & 0.1309\\
 red & 0.4629 & 0.0000 & 0.0000 & 0.2443 \\
 dark blue & 0.4670 & 0.0000 & 0.0000 & 0.3316\\
 turquoise & 0.4716 & 0.0000 & 0.0000 & 0.4014\\
 blue & 0.4887 & 0.0000 & 0.0000 & 0.5760 \\
 gray & 0.5211 & 0.0000 & 0.0000 & 0.7941\\
\hline\hline
\multicolumn{5}{c}{$R = 0.5000$}\\
\hline
 black & 0.3783 & 0.0000 & 0.0000 & 0.0175\\ 
 violet & 0.4053 & 0.0000 & 0.0000 & 0.4014\\ 
 red & 0.4607 & 0.0000 & 0.0000 & 0.7156 \\
 dark blue & 0.5016 & 0.0000 & 0.0000 & 0.8988 \\
\hline\hline
\multicolumn{5}{c}{$R = 0.4000$}\\
\hline
 black & 0.4413 & -1.5708 & 0.0000 & 0.3403\\ 
 violet & 0.3368 & -1.5708 & 0.0000 & 0.2705\\ 
 red & 0.2432 & -0.7854 & 0.0000 & 0.3054 \\
 dark blue & 0.0023 & 0.0000 & 0.0000 & 0.6458 \\
 turquoise & 0.1348 & 0.0000 & 0.0000 & 0.6021 \\
 blue & 0.2600 & 0.0000 & 0.0000 & 0.7069 \\
\hline \hline
\multicolumn{5}{c}{$R = 0.3000$}\\
\hline
 black & 0.0096 & 0.0000 & 0.0000 & 1.1606\\ 
 violet & 0.0267 & 0.0000 & 0.0000 & 1.1519\\ 
 red & 0.0051 & -1.5708 & 0.0000 & 0.3665 \\
 dark blue & 0.2154 & -0.7854 & 0.0000 & 0.5149 \\
 turquoise & 0.3407 & -0.7854 & 0.0000 & 0.5672 \\
\hline \hline
\multicolumn{5}{c}{$R = 0.2650$}\\
\hline
 black & 0.0088 & 0.0000 & 0.0000 & 1.4399\\ 
 violet & 0.0239 & 0.0000 & 0.0000 & 1.4224\\ 
 red & 0.1103 & 0.0000 & 0.0000 & 1.3875 \\
 dark blue & 0.2154 & 0.0000 & 0.0000 & 1.4748 \\
 turquoise & 0.2098 & -0.7854 & 0.0000 & 0.5760 \\
\hline \hline
\end{tabular}
\end{table}

\begin{table}
\caption{Initial conditions for Figure \ref{fig:nbody} ($\varpi_1=0$).}\label{tab:nbody}
\begin{tabular}{|r|r|r|r|r|} 
\hline\hline 
\multicolumn{1}{c}{$e_1$} & \multicolumn{1}{c}{$e_2$} & \multicolumn{1}{c}{$\Omega_2$} & \multicolumn{1}{c}{$\varpi_2$} & \multicolumn{1}{c}{$I$} \\
 & & \multicolumn{1}{c}{(rad)} & \multicolumn{1}{c}{(rad)} & \multicolumn{1}{c}{(rad)} \\
\hline\hline 
\multicolumn{5}{c}{Panel (a) $\omega_2$ resonance}\\
\hline 
 0.1 & 0.3000 & -1.5708 & 0.0000 & 1.0917 \\
\hline\hline 
\multicolumn{5}{c}{Panel (b) $\Omega_2-\varpi_1$ resonance}\\
\hline 
 0.7 & 0.5250 & 2.3562 & 2.3562 & 1.0137 \\ 
\hline\hline 
\multicolumn{5}{c}{Panel (c) $\varpi_2-\varpi_1$ resonance}\\
\hline 
0.7 & 0.2000 & 0.0000 & 0.0000 & 0.6500 \\
\hline \hline 
\multicolumn{5}{c}{Panel (d) $\varpi_2+\varpi_1-2\Omega_2$ resonance}\\
\hline 
 0.1 & 0.0787 & 0.0000 & 0.0000 & 1.3177\\ 
\hline \hline 
\multicolumn{5}{c}{Panel (e) $\varpi_2-3\varpi_1 + 2\Omega_2$ resonance}\\
\hline 
 0.7 & 0.7000 & 0.0000 & 0.0000 & 1.3000 \\ 
\hline \hline 
\end{tabular}
\end{table}


\bsp	
\label{lastpage}
\end{document}